\newcommand{\diracslash}[1]{#1\llap{/\kern2pt}}

\newcommand{\be}{\begin{equation}}
\newcommand{\ee}{\end{equation}}
\newcommand{\bea}{\begin{eqnarray}}
\newcommand{\eea}{\end{eqnarray}}
\newcommand{\ba}[1]{\begin{array}{#1}}
\newcommand{\ea}{\end{array}}

\newcommand{\bt}{\begin{tabular}}
\newcommand{\et}{\end{tabular}}

\newcommand{\beas}{\begin{eqnarray*}}
\newcommand{\eeas}{\end{eqnarray*}}

\documentclass[preprint,prd,aps,floats,nofootinbib,floatfix]{revtex4}

\DeclareSymbolFont{rsfs}{U}{rsfs}{m}{n}
\DeclareSymbolFontAlphabet{\mathrsfs}{rsfs}

\usepackage{graphicx}
\usepackage{multirow}
\usepackage{graphicx,epstopdf}

\usepackage{graphicx}

\usepackage[utf8]{inputenc}
\usepackage[english]{babel}
\usepackage{hyperref}
\hypersetup{
    colorlinks=true,
    linkcolor=blue,
    filecolor=magenta,     
    citecolor=blue, 
    urlcolor=blue,
}
 
\usepackage{cleveref}
\begin{document}

\title{  Quark Matter within Polyakov Chiral SU(3) Quark Mean Field Model at Finite Temperature} 
 \author{Manisha Kumari}
\email{manishak.phy.18@nitj.ac.in}
\affiliation{Department of Physics, Dr. B R Ambedkar National Institute of Technology Jalandhar, 
 Jalandhar -- 144011, Punjab, India}
\author{Arvind Kumar}
\email{iitd.arvind@gmail.com, kumara@nitj.ac.in}
\affiliation{Department of Physics, Dr. B R Ambedkar National Institute of Technology Jalandhar, 
 Jalandhar -- 144011, Punjab, India}

\def\be{\begin{equation}}
\def\ee{\end{equation}}
\def\bearr{\begin{eqnarray}}
\def\eearr{\end{eqnarray}}
\def\zbf#1{{\bf {#1}}}
\def\bfm#1{\mbox{\boldmath $#1$}}
\def\hf{\frac{1}{2}}
\def\kp{\zbf k+\frac{\zbf q}{2}}
\def\km{-\zbf k+\frac{\zbf q}{2}}
\def\hwo{\hat\omega_1}
\def\hwt{\hat\omega_2}

\begin{abstract}
 Thermodynamical properties of asymmetric strange quark matter using the Polyakov Chiral $\text{SU(3)}$ quark mean field (PCQMF) model at finite temperature and chemical potential have been investigated. Within the PCQMF model, 
the properties of quark matter are calculated through the 
scalar fields $\sigma$, $\zeta$, $\delta$ and  $\chi$, the vector fields $\omega$, $\rho$ and $\phi$ and the Polyakov loop fields $\Phi$ and $\bar{\Phi}$. 
 The isospin splitting of constituent quark masses is observed at large isospin asymmetry. The effect of temperature and strangeness fraction on energy per baryon and equation of state is found to be appreciable in quark matter. The effect of the Polyakov loop dynamics on several thermodynamical bulk quantities such as energy density, entropy density, and trace anomaly is presented and compared with recent lattice QCD results. 
\end{abstract}

\maketitle

\maketitle

\section{Introduction}
\label{intro}

The thermodynamical properties of hot and dense matter under extreme conditions of temperature and density may play an important role to probe the physics shortly after the big bang \cite{Boyanovsky2006, Boeckel2011}, outcome of heavy-ion collisions \cite{Marty2013}, properties of supernova explosions \cite{Janka2012} and the structure of compact stars \cite{Weber2005}. The phase diagram of QCD represents the information about the equilibrium phases of QCD as well as the physics of phase transitions in the plane of baryon chemical potential, $\mu_B$, and temperature $T$. Subjecting hadronic matter to high temperature and/or density may result in the restoration of chiral symmetry and transition to the QGP/quark matter phase. 
A quark matter phase at low temperature and at large baryonic chemical potential, $\mu_B$ (in which quark Cooper pairs are formed) may be a color superconducting phase \cite{alford2001, Rischke2004}. The Critical-End-Point (CEP) at which the chiral phase transition changes its behavior, is a fundamental landmark of the QCD phase diagram, but its exact location is still an open issue \cite{Fodor2004,Iwasaki2004}. A new phase of QCD matter, $i.e.$, quarkyonic phase was proposed at high baryon number density \cite{McLerran2007,McLerran2009}. The matter that exists in the quarkyonic phase is expected to have energy density and pressure as that of gas of confined quarks \cite{McLerran2009,Fukushima2004,Ratti2006,Pisarski2000,Fukushima2008}.
It has been considered that the Strange Quark Matter (SQM) exists in quark stars \cite{menezes2006,Kalam2013} as well as in the core of neutron stars \cite{Lastowiecki2015,Annala2019}. If the energy per baryon for the SQM becomes less than the stable atomic nuclei (Fe and Ni), than SQM can be a true ground state of matter \cite{Bodmer1971,Farhi1984,Witten1984}.  The possible existence of strange stars (made entirely of deconfined $u$, $d$ and $s$ quark matter or strange matter) is one of the most exciting issue of modern physics \cite{Thirukkanesh2017, Shahzad2019}.

The experimental programs, for example, Relativistic Heavy Ion Collider (RHIC) at BNL and the Large Hadron Collider (LHC) at CERN \cite{ramona2007,helios1995,agaki1995,porter1997,wilson1998,sashi2011,Hbook1994} help to understand the QCD matter at high temperatures and almost zero baryonic density. Future facilities such as Nuclotron-based Ion Collider Facility (NICA) at JINR Dubna \cite{sissa2009,keke2012}, Japan Proton Accelerator Research Complex (J-PARC) and the Facility for Antiproton and Ion Research (FAIR) at GSI Germany are being constructed to study the regime of high baryonic density and moderate temperature with a goal to understand the position of CEP, phase boundaries, and properties of quark matter.

Alongside the experimental endeavors, there are many theoretical frameworks used to study the properties of hot and dense matter. These are, for example, Dyson-Schwinger equation approach \cite{Roberts2000,Alkofer2001,Maris2003,Xu2015}, Quark Mass Density Dependent (QMDD) model \cite{Fowler1981,Chakrabarty1989,Chakrabarty1991,Chakrabarty1993,Benvenuto1995}, Quark-Meson Coupling (QMC) model \cite{Tsushima1998}, Polyakov-Quark Meson Coupling (PQMC) model \cite{Schaefer2007,Stiele2014}, Nambu-Jona-Lasinio (NJL) model (with and without four-vector-type interactions) \cite{peng2017,menezes2006}, Polyakov extended NJL (PNJL) model \cite{Costa2010,Sakai1991,Sasaki1991}, the Entanglement PNJL (EPNJL) model \cite{Sakai2010,Sasaki2012,Restrepo2015,Gatto2011}, Confined  Density Dependent Quark Mass (CDDM) model \cite{Peng2000,Peng2008,Rajagopal2011}  and Chiral  $\text{SU(3)}$ Quark Mean Field (CQMF) model \cite{wang2001,wang2001a}.

 The primal analysis of theoretical models that are used to investigate the properties of quark matter is based on the MIT bag model in which quarks are supposed to be confined within a phenomenological bag \cite{Weber2005}. 
The model calculation suggests that the SQM is absolutely stable in a range of parameters. Chin and Kerman conjectured that the SQM with $A\leq10$ might be metastable with half-life $\leq10^{-4}s$ \cite{Chin1979}. Moreover, Jaffe and Berger explained the surface correction for the strangelets where they predicted that the surface tension would destabilize strangelets \cite{Berger1987}. It has been hypothesized that there might exist compact astrophysical objects composed entirely of strange matter called strange stars \cite{Xu2008}. Firstly, with the help of the QMDD model properties of quark matter \cite{Chakrabarty1989} were investigated and then employed to study SQM \cite{Benvenuto1995,Zhang2001}. The QMDD model can give a dynamical explanation of confinement and stability and many thermodynamical properties of SQM at zero and finite temperature. But this model failed to explain the phase transitions of quark deconfinement because  quark masses are considered as independent of temperature \cite{Zhang2002}. To overcome this difficulty, one needs a Quark Mass Density and Temperature-Dependent (QMDTD) model in which quark confinement is temporary \cite{Qian2005}.

 The NJL model has been used for interpreting hadron properties \cite{Bentz2001}, phase transition \cite{wang2007,Schwarz1999} and multi-particle bound states \cite{Buballa1999}.
The introduction of the scalar-isovector and vector-isovector coupling within the NJL model is significant to study the effect of isospin on the quark matter \cite{Liu2019}. In the QMC model, the interaction between hadrons is mediated by the exchange of scalar and vector mesons self-consistently coupled to the quarks within those hadrons. The NJL or QMC models are undoubtedly an effective tool for interpretation of chiral symmetry breaking but does not explain deconfinement. To avoid this difficulty,  Polyakov loop potential coupled with NJL or QMC models are introduced. 
  The PNJL model which includes both the chiral dynamics and deconfinement effect at extreme temperature provides a good description of lattice data at zero chemical potential \cite{Ratti2006,Costa2010}.

 In the present investigation to study the properties of isospin asymmetric quark matter at finite temperature and density, with finite strangeness chemical potential, we will couple Polyakov potential with Chiral  $\text{SU(3)}$ Quark Mean Field (CQMF) model. Henceforth, we will name this as the PCQMF model.   
  In the CQMF model, quarks are confined within baryons by an effective potential. This model has been applied to investigate nuclear matter \cite{wang2001}, strange hadronic matter \cite{wang2001a,Harpreet2018}, finite nuclei, hypernuclei \cite{wang2002}. The original model was improved by using the linear definition of effective baryon mass \cite{wang2004a}.  The strange quark matter properties have been studied in the CQMF model at zero temperature \cite{wang2003}.
   CQMF model can also explain the hadronic and quark matter phase and can further be used for the study of a mixed-phase, where both the hadrons and quarks can exist simultaneously \cite{wang2007}.
 This model has also been used to investigate the in-medium magnetic moments of octet and decuplet baryons \cite{Harpreet2018,Harpreet2018a}.  
     In the present model, the interaction of quarks is explained in terms of non-strange scalar meson field, $\sigma$, strange scalar meson field, $\zeta$, scalar isovector meson field, $\delta$, scalar isoscalar dilaton field, $\chi$, non-strange vector field, $\omega$, non-strange vector-isovector  field, $\rho$ and strange vector field, $\phi$. In this model, a mean-field approximation is applied, which is based on the non-perturbative relativistic approach and generally used for the explanation of the many-body interaction. In this approximation, mesons are analyzed as classical fields, hence the quark-meson interaction Lagrangian term includes  only the contribution from scalar and vector fields.

Effective chiral models, associated with any form of the Polyakov loop potential, usually have their parameters modified to give good reproduction of lattice data at vanishing density. Testing the extended effective model is critical to explore whether these models provide a quantitative and qualitative rigorous explanation of the strongly interacting matter. By introducing the Polyakov loop potential, these models have become a very attractive approach to incorporate the hadronic and quark matter and involvement of gluonic degrees of freedom. In this model, a supplementary Polyakov potential must be proposed to explain the deconfinement transition and incorporate the thermal fluctuations in the pure gluonic theory. One can also include the thermal fluctuation by using Functional Renormalization Group (FRG) techniques \cite{Herbst2014,Drews2013}.

The present paper is organized as follows: In Section II (A), we will briefly describe the chiral $\text{SU(3)}$ quark mean-field model and derive the grand canonical potential density by applying mean-field approximation. In Section II (B), we will introduce the Polyakov loop variable and discuss its effective potential. The Polyakov loop potential is then coupled to the chiral $\text{SU(3)}$ quark mean-field model. With the help of modified thermodynamical potential density, we have calculated different thermodynamical quantities. In Section III, we will explain the result of our analysis on asymmetric quark matter and finally, in Section IV the results of present work will be summarized.

\section{Methodology}
\label{sec:2}                                                                   \subsection{ Chiral $\text{SU(3)}$ Quark Mean Field (CQMF) Model} 
                                                                   \label{sec:2a}
 In the CQMF model, we describe the quark-meson and meson-meson interactions based on broken scale invariance \cite{Papazoglou1999,Mishra2004,Mishra2004a} and non-linear realization of chiral $\text{SU(3)}$ symmetry \cite{Weinberg1968,Coleman1969,Bardeen1969} at finite temperature and density. The masses of quark and mesons (except pseudoscalar meson) are obtained by the mechanism of spontaneous symmetry breaking and pseudoscalar meson get their masses through explicitly symmetry breaking term which satisfy the partially conserved axial-vector current (PCAC) relation. In Ref. \cite{wang2001,wang2001a} where CQMF was used to investigate the properties of nuclear and strange hadronic matter, quarks are confined inside hadrons through a confined potential. However, in the present work, as we are interested in the properties of quark matter only, we need not to consider such kind of potential \cite{wang2003}.
The total effective Lagrangian density in chiral $\text{SU(3)}$ quark mean field model to describe the SQM is given by 
\begin{equation}
{\cal L}_{{\rm eff}} \, = \, {\cal L}_{q0} \, + \, {\cal L}_{qm}
\, + \,
{\cal L}_{\Sigma\Sigma} \,+\, {\cal L}_{VV} \,+\, {\cal L}_{SB}\,
+ \, {\cal L}_{\Delta m} \, + \, {\cal L}_{h}. \label{totallag}
\end{equation}
  
In the above equation, ${\cal L}_{q0} = \bar q \, i\gamma^\mu \partial_\mu q $ is the free part of massless quarks, ${\cal L}_{qm}$ represents the quark mesons interaction term which is invariant under the chiral  $\text{SU(3)}$ transformation and is given by 
\begin{eqnarray}
{\cal L}_{qm}  =  g_s\left(\bar{q}_LM q_R+\bar{q}_RM^+q_L\right)- g_v\left(\bar{q}_L\gamma^\mu l_\mu q_L+\bar{q}_R\gamma^\mu r_\mu q_R\right),  \nonumber \\
\end{eqnarray}

where $q=\left(\begin{array}{lcr}
u \\
d \\
s
\end{array}\right)$ and $g_v$ and $g_s$ are vector and scalar coupling constants, respectively. The compact form of spin-0 scalar ($\Sigma$) and pseudoscalar ($\Pi$) meson nonets  can be expressed as 
\begin{equation}
M(M^{\dagger})=\Sigma \pm i\Pi =\frac{1}{\sqrt{2}}\sum_{a=0}^{8}
\left( \sigma^{a}\pm i \pi ^{a}\right) \lambda ^{a},
\end{equation}
where $\sigma^{a}$ and $\pi^{a}$ represent the nonets of scalar and pseudoscalar mesons, respectively,  $\lambda ^{a}$ are Gell-Mann matrices with $\lambda ^{0}=\sqrt{\frac{2}{3}}I$. 
We can define spin-1 meson in a similar way as 
\begin{equation}
l_{\mu }(r_{\mu })=\frac{1}{2}\left( V_{\mu }\pm A_{\mu }\right)
= \frac{1}{2\sqrt{2}}\sum_{a=0}^{8}\left( v_{\mu}^{a}\pm a^{a}_{\mu}
\right) \lambda^{a}.
\end{equation}
In above, $v_{\mu}^a$ and $a_{\mu}^a$ are nonets of vector and pseudovector mesons, respectively.
 The expressions for physical states of scalar and vector meson nonets  are 
\begin{equation}
\Sigma = \frac1{\sqrt{2}}\sum_{a=0}^8 \sigma^a \, \lambda^a=\left(
\begin{array}{lcr}
\frac1{\sqrt{2}}\left(\sigma+\delta^0\right) & \delta^{+} & \kappa^{*+} \\
\delta^- & \frac1{\sqrt{2}}\left(\sigma-\delta^0\right) & \kappa^{*0} \\
\kappa^{*-} & \bar{\kappa}^{*0} & \zeta
\end{array}
\right),
\end{equation}
and
\begin{equation}
V_\mu = \frac1{\sqrt{2}}\sum_{a=0}^8 v_\mu^a \, \lambda^a=\left(
\begin{array}{lcr}
\frac1{\sqrt{2}}\left(\omega_\mu+\rho_\mu^0\right)
& \rho_\mu^+ & K_\mu^{*+}\\
\rho_\mu^- & \frac1{\sqrt{2}}\left(\omega_\mu-\rho_\mu^0\right)
& K_\mu^{*0}\\
K_\mu^{*-} & \bar{K}_\mu^{*0} & \phi_\mu
\end{array}
\right),
\,
\end{equation}
respectively.

In the mean-field approximation, the chiral-invariant scalar meson self-interaction term, ${\cal L}_{\Sigma\Sigma}$ ($3^{\text{rd}}$ term in \cref{totallag}), and vector meson self-interaction term, ${\cal L}_{VV}$ ($4^{\text{th}}$ term in \cref{totallag}), are written as

\begin{eqnarray}
{\cal L}_{\Sigma\Sigma} =& -\frac{1}{2} \, k_0\chi^2
\left(\sigma^2+\zeta^2+\delta^2\right)+k_1 \left(\sigma^2+\zeta^2+\delta^2\right)^2
+k_2\left(\frac{\sigma^4}{2} +\frac{\delta^4}{2}+3\sigma^2\delta^2+\zeta^4\right)\nonumber \\ 
&+k_3\chi\left(\sigma^2-\delta^2\right)\zeta 
 -k_4\chi^4-\frac14\chi^4 {\rm ln}\frac{\chi^4}{\chi_0^4} +
\frac{d}
3\chi^4 {\rm ln}\left(\left(\frac{\left(\sigma^2-\delta^2\right)\zeta}{\sigma_0^2\zeta_0}\right)\left(\frac{\chi^3}{\chi_0^3}\right)\right), \label{scalar0}
\end{eqnarray}    
and
\begin{equation}
{\cal L}_{VV}=\frac{1}{2} \, \frac{\chi^2}{\chi_0^2} \left(
m_\omega^2\omega^2+m_\rho^2\rho^2+m_\phi^2\phi^2\right)+g_4\left(\omega^4+6\omega^2\rho^2+\rho^4+2\phi^4\right), \label{vector}
\end{equation}

respectively. Here $d$ =6/33 for three colors and three flavors and $\sigma_0$, $\zeta_0$ and $\chi_0$ are the vacuum expectation values of the corresponding $\sigma$, $\zeta$ and $\chi$ fields. 
In \cref{vector} vector meson masses are density-dependent which can be expressed as
 
\begin{eqnarray}
m^2_\omega = m_\rho^2 =
\frac{m_v^2}{1 - \frac{1}{2} \mu \sigma^2}\, ,
\hspace*{.5cm} {\rm and} \hspace*{.5cm}
m^2_\phi = \frac{m_v^2}{1 - \mu \zeta^2}\, .
\end{eqnarray}
In above equation, the vacuum value of the vector meson mass $m_v$ = 673.6 MeV and density parameter $\mu$ = 2.34 fm$^2$ are taken to reproduce ${m_\omega}$ = 783 MeV and ${m_\phi}$ = 1020 MeV.
The last three terms ${\cal L}_{SB}$, ${\cal L}_{\Delta m}$ and ${\cal L}_{h}$ of \cref{totallag} breaks the chiral symmetry explicitly. The relations among various quark meson coupling constant as required by chiral symmetry as defined through relations 
\begin{eqnarray}
\frac{g_s}{\sqrt{2}} 
&=& g_{\delta}^u = -g_{\delta}^d = g_\sigma^u = g_\sigma^d = \ldots = 
\frac{1}{\sqrt{2}}g_\zeta^s,
~~~~~g_{\delta}^s = g_\sigma^s = g_\zeta^u = g_\zeta^d = 0 \, ,\\
\frac{g_v}{2\sqrt{2}} 
&=& g_{\rho^0}^u = -g_{\rho^0}^d = g_\omega^u = g_\omega^d = \ldots = 
\frac{1}{\sqrt{2}}g_\phi^s,
~~~g_\omega^s = g_{\rho^0}^s = g_\phi^u = g_\phi^d = 0.
\end{eqnarray}    
The Lagrangian density ${\cal L}_{SB}$ in \cref{totallag} generates non-vanishing masses for pseudoscalar mesons and is given as
\begin{equation} 
{\cal L}_{SB}=-\frac{\chi^2}{\chi_0^2}\left[m_\pi^2f_\pi\sigma + 
\left(\sqrt{2}m_K^2f_K-\frac{m_\pi^2}{\sqrt{2}}f_\pi\right)\zeta\right].
\end{equation} 
This lead to a non-vanishing divergence of the axial currents which satisfy the Partial Conserved Axial-vector Current  (PCAC) relations for $\pi$ and $K$ mesons (masses of $\pi$ and $K$ mesons are not zero).
The parameters  $\sigma_0$ and  $\zeta_0$ are constrained by the spontaneous  breaking of chiral symmetry and are expressed in terms of pion decay constant ($f_\pi$ = 93 MeV) and the kaon decay constant ($f_K$ = 115 MeV) as: 
\begin{eqnarray}
\sigma_0 = - f_\pi    \hspace*{1cm} \hspace*{.5cm} {\rm and} \hspace*{.8cm}
\zeta_0  = \frac{1}{\sqrt{2}} ( f_\pi - 2 f_K). 
\end{eqnarray}

  In order to obtain exact constituent mass of strange quark, we need to include an additional mass term
\begin{eqnarray} 
{\cal L}_{\Delta m} = - \Delta m_s \bar q S q,
\end{eqnarray}
where $S \, = \, \frac{1}{3} \, \left(I - \lambda_8\sqrt{3}\right) = 
{\rm diag}(0,0,1)$ define the strangeness quark matrix and $\Delta m_s = 29$ MeV . The relations for constituent quark masses in vacuum are written as
\begin{eqnarray}
m_u=m_d=-\frac{g_s}{\sqrt{2}}\sigma_0, ~~~ 
m_s=-g_s \zeta_0 + \Delta m_s. 
\end{eqnarray}
To obtain reasonable hyperon potential, we should add an additional symmetry breaking term. It can be expressed in the presence of mean field approximation as \cite{wang2001a}
\begin{eqnarray}
{\cal L}_h \, = \, (h_1 \, \sigma \,  + \, h_2 \, \zeta) \, \bar{s} s \, .  
\end{eqnarray}

 Including the mean field approximation and some finite temperature field theory algebra, the thermodynamical potential density of SQM at finite temperature and density can be written as
\begin{eqnarray}
\hspace*{-.4cm} 
\Omega_{\rm{CQMF}}= {\Omega}_{q\bar{q}}- {\cal L}_M- {\cal V}_{vac},
\end{eqnarray}
where ${\Omega}_{q\bar{q}}$ represents the contribution of  quarks and antiquarks to the total thermodynamical potential and defined by
\begin{eqnarray}
\hspace*{-.4cm} 
{\Omega}_{q\bar{q}}=\sum_{i=u,d,s}\frac{-\gamma_i k_BT N_c}{(2\pi)^3}\int_0^\infty 
d^3k\left\{ {\rm ln} 
\left(1+e^{-(E_i^*(k)-{\nu_i}^{*})/k_BT}\right)+
{\rm ln} \left(1+e^{-(E_i^*(k)+{\nu_i}^{*})/k_BT}\right)\right\},
\end{eqnarray}

where summation runs over constituent quarks. Also, $\gamma_i$ = 2 is spin degeneracy factor, $N_c$= 3 is the color degree of freedom and $E_i^*(k)=\sqrt{m_i^{*2}+k^2}$  is the effective single particle energy of quarks. The term ${\cal L}_M = {\cal L}_{\Sigma\Sigma} +{\cal L}_{VV} +{\cal L}_{SB}$ describe the interaction between mesons. The vacuum energy term, ${\cal V}_{vac}$ is subtracted in order to get zero vacuum energy. The effective chemical potential ${\nu_i}^{*}$ of quarks is related to the usual chemical potential $\mu_i$ by
\begin{equation}
{\nu_i}^{*}=\mu_i-g_\omega^i\omega-g_\phi^i\phi-g_\rho^i\rho,
\label{mueff}
\end{equation} 
where  $g^i_{\omega}$, $g^i_{\phi}$ and $g^i_{\rho}$ are the coupling strength of quarks with vector meson fields.

The effective constituent quark mass ${m_i}^{*}$ is defined by the relation   
\begin{equation}
{m_i}^{*} = -g_{\sigma}^i\sigma - g_{\zeta}^i\zeta - g_{\delta}^i\delta + {m_{i0}}.
\label{mbeff}
\end{equation}
In the above equation, $g_{\sigma}^i$, $g_{\zeta}^i$ and $g_{\delta}^i$ represent the coupling strengths of different quarks with scalar fields. The values of  $g_{\sigma}^i$, $g_{\zeta}^i$ and  ${m_{i0}}$ are chosen to fit the vacuum masses of constituent quarks which are taken to be  $m_u= m_d=313$ MeV and $m_s=490$ MeV \cite{wang2003}.

The number (vector) density, $\rho_{i}$, and scalar density, $\rho_{i}^{s}$, of quarks is defined as 
\begin{eqnarray}
\rho_{i} = \gamma_{i}N_c\int\frac{d^{3}k}{(2\pi)^{3}}  
\Big(f_i(k)-\bar{f}_i(k)
\Big),
\label{rhov0}
\end{eqnarray}
 and
\begin{eqnarray}
\rho_{i}^{s} = \gamma_{i}N_c\int\frac{d^{3}k}{(2\pi)^{3}} 
\frac{m_{i}^{*}}{E^{\ast}_i(k)} \Big(f_i(k)+\bar{f}_i(k)
\Big),
\label{rhos0}
\end{eqnarray}
respectively, where $f_i(k)$ and $\bar{f}_i(k)$ represent the Fermi distribution functions at finite temperature for quarks and anti-quarks and are expressed as 
\bea
f_i(k) &=& \frac{1}{1+\exp\left[(E^*_{i}(k) 
-\nu^{*}_{i})/k_BT \right]}~~~ \text{and}~~~
\bar{f}_i(k) = \frac{1}{1+\exp\left[( E^*_{i}(k) 
+\nu^{*}_{i})/k_BT\right]}~.
\label{dfp}
\eea

\subsection{Polyakov Chiral $\text{SU(3)}$ Quark Mean Field Model }
\label{sec:2b}

The boundary conditions in pure glue theory (quark having infinite mass) are respected by the $Z(N_C) $ symmetry. For $Z(N_C) $ symmetry breaking, an order parameter in terms of thermal Wilson line is defined as  \cite{shao2016}, 
\begin{equation}
   L(\vec{x})=\mathcal{P} exp\bigg[i\int_0^{\frac{1}{T}} d\tau A_4 (\vec{x},\tau)   \bigg].
\end{equation}

In above, $\mathcal{P}$ is the path ordering operator and $A_4$ is gluon field in temporal direction which is defined as \cite{shao2016}
\begin{equation}
A_\mu=ig_sA_\mu^{a}\frac{\lambda_a}{2}\delta_0^\mu, ~~~~~~ a=1,....{N^{2}_C}-1,
\end{equation}

here, $A_\mu^{a}$ represents the gluon field of color index $a$.

The Polyakov loop variable, $\Phi(\vec{x})$ and its conjugate, $\bar{\Phi}(\vec{x})$ can be defined as the thermal expectation value of trace over color of the thermal Wilson line, i.e., \cite{Polyakov1978}
\begin{equation}
\Phi(\vec{x})=(Tr_c L)/N_C,
~~~~~~ \bar{\Phi}(\vec{x})=(Tr_c L^\dag)/N_C.
\end{equation}

 By introducing static gluonic degrees of freedom in the CQMF model through an effective gluon potential in terms of Polyakov loop, it is possible to study features of both chiral symmetry breaking and deconfinement in the improved CQMF model named as Polyakov  Chiral $\text{SU(3)}$ Quark Mean Field (PCQMF) model. 
 
 The total Lagrangian density of SQM in PCQMF model is modified as
\begin{equation}
{\cal L}_{{\rm PCQMF}} \, = \, {\cal L}_{\rm eff} \, -U(\Phi(\vec{x}),\bar{\Phi}(\vec{x}),T), \label{PCQMFlag}
\end{equation}

where $U(\Phi(\vec{x}),\bar{\Phi}(\vec{x}),T)$ is temperature dependent Polyakov loop effective potential. In this work, for Polyakov potential we will consider the commonly used Logarithmic form, \cite{Costa2010,Fukushima2004,Roessner2007} which satisfies the $Z(N_C) $ symmetry of pure gauge Lagrangian and is given by
\begin{eqnarray}
     \frac{U(\Phi,\bar{\Phi},T)}{T^4}&=&-\frac{a(T)}{2}\bar{\Phi}\Phi+b(T)\mathrm{ln}\big[1-6\bar{\Phi}\Phi+4(\bar{\Phi}^3+\Phi^3)-3(\bar{\Phi}\Phi)^2\big].
\end{eqnarray}

The $T$-dependent parameters $a(T)$ and $b(T)$ appearing in the above equation are given by \cite{Costa2010,Roessner2007}:

\begin{equation}\label{T}
    a(T)=a_0+a_1\bigg(\frac{T_0}{T}\bigg)+a_2\bigg(\frac{T_0}{T}\bigg)^2,\ \  b(T)=b_3\bigg(\frac{T_0}{T}\bigg)^3.
\end{equation}

The parameters $a_0$, $a_1$, $a_2$ and $b_3$ summarized in Table \ref{tab:1} are
precisely fitted according to the result of lattice QCD thermodynamics in
pure gauge sector \cite{Roessner2007}. 

\begin{table}[h]
\centering
\begin{tabular}{|l|l|l|l|}
\hline
\multicolumn{1}{|c|}{$a_0$} & \multicolumn{1}{c|}{$a_1$} & \multicolumn{1}{c|}{$a_2$} & \multicolumn{1}{c|}{$b_3$} \\ \hline
1.81                        & -2.47                      & 15.2                       & -1.75                      \\ \hline
\end{tabular}
\caption{Parameters in Polyakov effective potential}
\label{tab:1}
\end{table}

The parameter $T_0=270$ MeV is the confinement-deconfinement transition temperature in the pure Yang-Mills theory at vanishing chemical potential \cite{Fukugita1990}. The effective potential reveals the aspect of phase transition from confinement ($T<T_0$, the minima of the potential being at $\Phi=0$) to deconfinement ($T>T_0$, the minima of the potential being at $\Phi\neq0$) \cite{Costa2010}.

The thermodynamical potential density of SQM in the PCQMF model at finite temperature and density within the mean field approximation can be expressed as
\begin{eqnarray}\label{omegapnjl}
&&\Omega=\mathcal{U}(\Phi,\bar{\Phi},T)
+\Omega_{q\bar{q}}-{\cal L}_M- {\cal V}_{vac}.
\end{eqnarray}
In above, quark and antiquark thermal contribution, $\Omega_{q\bar{q}}$, is modified to:
 \begin{eqnarray}\label{omegaq}
\Omega_{q\bar{q}}=-\gamma_i k_BT\sum_{i=u,d,s}\int_0^\infty\frac{d^3k}{(2\pi)^3}[\ln(1+e^{-3(E_i^*(k)-{\nu_i}^{*})/k_BT}+3\Phi e^{-(E_i^*(k)-{\nu_i}^{*})/k_BT}
\nonumber\\
+3\bar{\Phi}e^{-2(E_i^*(k)-{\nu_i}^{*})/k_BT})+\ln(1+e^{-3(E_i^*(k)+{\nu_i}^{*})/k_BT}+
\nonumber\\
3\bar{\Phi} e^{-(E_i^*(k)+{\nu_i}^{*})/k_BT}
+3\Phi e^{-2(E_i^*(k)+{\nu_i}^{*})/k_BT})].
\end{eqnarray}

By including the Polyakov loop potential, fermion distribution functions of quarks, $f_{i}(k)$ and antiquarks, $\bar{f}_{i}(k)$, also modifies to
\begin{equation}\label{distribution}
  f_{i}(k)=\frac{\Phi e^{-(E_i^*(k)-{\nu_i}^{*})/k_BT}+2\bar{\Phi} e^{-2(E_i^*(k)-{\nu_i}^{*})/k_BT}+e^{-3(E_i^*(k)-{\nu_i}^{*})/k_BT}}
  {1+3\Phi e^{-(E_i^*(k)-{\nu_i}^{*})/k_BT}+3\bar{\Phi} e^{-2(E_i^*(k)-{\nu_i}^{*})/k_BT}+e^{-3(E_i^*(k)-{\nu_i}^{*})/k_BT}} , 
\end{equation}

\begin{equation}\label{distribution1}
  \bar{f}_{i}(k)=\frac{\bar{\Phi} e^{-(E_i^*(k)+{\nu_i}^{*})/k_BT}+2{\Phi} e^{-2(E_i^*(k)+{\nu_i}^{*})/k_BT}+e^
  {-3(E_i^*(k)+{\nu_i}^{*})/k_BT}}{1+3\bar{\Phi} e^{-(E_i^*(k)+{\nu_i}^{*})/k_BT}+3{\Phi} e^{-2(E_i^*(k)+{\nu_i}^{*})/k_BT}+e^{-3(E_i^*(k)+{\nu_i}^{*})/k_BT}} .
\end{equation}

To determine the scalar fields $\sigma$, $\zeta$ and $\delta$, the dilaton field $\chi$, the vector fields $\omega$, $\rho$ and $\phi$ and the Polyakov field $\Phi$ and its conjugate $\bar{\Phi}$, we minimize $\Omega$ with respect to these fields, $i.e.$,

\begin{equation}
\frac{\partial\Omega}
{\partial\sigma}=\frac{\partial\Omega}{\partial\zeta}=\frac{\partial\Omega}{\partial\delta}=\frac{\partial\Omega}{\partial\chi}=\frac{\partial\Omega}{\partial\omega}=\frac{\partial\Omega}{\partial\rho}=\frac{\partial\Omega}{\partial\phi}=\frac{\partial\Omega}{\partial\Phi}=\frac{\partial\Omega}{\partial\bar\Phi}=0.
\end{equation}
This results in following system of coupled equations:

 \begin{eqnarray}\label{sigma1}
&&\frac{\partial \Omega}{\partial \sigma}= k_{0}\chi^{2}\sigma-4k_{1}\left( \sigma^{2}+\zeta^{2}
+\delta^{2}\right)\sigma-2k_{2}\left( \sigma^{3}+3\sigma\delta^{2}\right)
-2k_{3}\chi\sigma\zeta \nonumber\\
&-&\frac{d}{3} \chi^{4} \bigg (\frac{2\sigma}{\sigma^{2}-\delta^{2}}\bigg )
+\left( \frac{\chi}{\chi_{0}}\right) ^{2}m_{\pi}^{2}f_{\pi}- 
\left(\frac{\chi}{\chi_0}\right)^2m_\omega\omega^2
\frac{\partial m_\omega}{\partial\sigma}\nonumber\\
 &-&\left(\frac{\chi}{\chi_0}\right)^2m_\rho\rho^2 
\frac{\partial m_\rho}{\partial\sigma}
-\sum_{i=u,d} g_{\sigma}^i\rho_{i}^{s} = 0 ,
\end{eqnarray}
\begin{eqnarray}
&&\frac{\partial \Omega}{\partial \zeta}= k_{0}\chi^{2}\zeta-4k_{1}\left( \sigma^{2}+\zeta^{2}+\delta^{2}\right)
\zeta-4k_{2}\zeta^{3}-k_{3}\chi\left( \sigma^{2}-\delta^{2}\right)-\frac{d}{3}\frac{\chi^{4}}{{\zeta}}\nonumber\\
&+&\left(\frac{\chi}{\chi_{0}} \right)
^{2}\left[ \sqrt{2}m_{K}^{2}f_{K}-\frac{1}{\sqrt{2}} m_{\pi}^{2}f_{\pi}\right]-\left(\frac{\chi}{\chi_0}\right)^2m_\phi\phi^2 
\frac{\partial m_\phi}{\partial\zeta}
 -\sum_{i=s} g_{\zeta}^i\rho_{i}^{s} = 0 ,
\label{zeta}
\end{eqnarray}
\begin{eqnarray}
&&\frac{\partial \Omega}{\partial \delta}=k_{0}\chi^{2}\delta-4k_{1}\left( \sigma^{2}+\zeta^{2}+\delta^{2}\right)
\delta-2k_{2}\left( \delta^{3}+3\sigma^{2}\delta\right) +\mathrm{2k_{3}\chi\delta
\zeta} \nonumber\\
& + &  \frac{2}{3} d \chi^4 \left( \frac{\delta}{\sigma^{2}-\delta^{2}}\right)
-\sum_{i=u,d} g_{\delta}^i\rho_{i}^{s} = 0 ,
\label{delta}
\end{eqnarray}
\begin{eqnarray}
&&\frac{\partial \Omega}{\partial \chi}=\mathrm{k_{0}\chi} \left( \sigma^{2}+\zeta^{2}+\delta^{2}\right)-k_{3}
\left( \sigma^{2}-\delta^{2}\right)\zeta + \chi^{3}\left[1
+{\rm {ln}}\left( \frac{\chi^{4}}{\chi_{0}^{4}}\right)  \right]
+(4k_{4}-d)\chi^{3}
\nonumber\\
&-&\frac{4}{3} d \chi^{3} {\rm {ln}} \Bigg ( \bigg (\frac{\left( \sigma^{2}
-\delta^{2}\right) \zeta}{\sigma_{0}^{2}\zeta_{0}} \bigg )
\bigg (\frac{\chi}{\mathrm{\chi_0}}\bigg)^3 \Bigg )+
\frac{2\chi}{\chi_{0}^{2}}\left[ m_{\pi}^{2}
f_{\pi}\sigma +\left(\sqrt{2}m_{K}^{2}f_{K}-\frac{1}{\sqrt{2}}
m_{\pi}^{2}f_{\pi} \right) \zeta\right] \nonumber\\
&-& \frac{\chi}{{\chi^2}_0}({m_{\omega}}^2 \omega^2+{m_{\rho}}^2\rho^2)  = 0,
\label{chi}
\end{eqnarray}
  \begin{eqnarray}
\frac{\partial \Omega}{\partial \omega}=\frac{\chi^2}{\chi_0^2}m_\omega^2\omega+4g_4\omega^3+12g_4\omega\rho^2
&-&\sum_{i=u,d}g_\omega^i\rho_{i}^{v}=0,
\label{omega} 
\end{eqnarray}
  \begin{eqnarray}
\frac{\partial \Omega}{\partial \rho}=\frac{\chi^2}{\chi_0^2}m_\rho^2\rho+4g_4\rho^3+12g_4\omega^2\rho&-&
\sum_{i=u,d}g_\rho^i\rho_{i}^{v}=0, 
\label{rho} 
\end{eqnarray}
  \begin{eqnarray}
\frac{\partial \Omega}{\partial \phi}=\frac{\chi^2}{\chi_0^2}m_\phi^2\phi+8g_4\phi^3&-&
\sum_{i=s}g_\phi^i\rho_{i}^{v}=0,
 \label{phi}  
\end{eqnarray}
 \begin{eqnarray}
\hspace*{0.4cm} 
\frac{\partial \Omega}{\partial \Phi} =\bigg[\frac{-a(T)\bar{\Phi}}{2}-\frac{6b(T)
(\bar{\Phi}-2{\Phi}^2+{\bar{\Phi}}^2\Phi)
}{1-6\bar{\Phi}\Phi+4(\bar{\Phi}^3+\Phi^3)-3(\bar{\Phi}\Phi)^2}\bigg]T^4
-\sum_{i=u,d,s}\frac{2k_BTN_C}{(2\pi)^3}
\nonumber\\
\int_0^\infty d^3k 
\bigg[\frac{e^{-(E_i^*(k)-{\nu_i}^{*})/k_BT}}{(1+e^{-3(E_i^*(k)-{\nu_i}^{*})/k_BT}+3\Phi e^{-(E_i^*(k)-{\nu_i}^{*})/k_BT}
+3\bar{\Phi}e^{-2(E_i^*(k)-{\nu_i}^{*})/k_BT})}
\nonumber\\
+\frac{e^{-2(E_i^*(k)+{\nu_i}^{*})/k_BT}}{(1+e^{-3(E_i^*(k)+{\nu_i}^{*})/k_BT}
+3\bar{\Phi} e^{-(E_i^*(k)+{\nu_i}^{*})/k_BT}+3\Phi e^{-2(E_i^*(k)+{\nu_i}^{*})/k_BT})}\bigg]=0,
\label{Polyakov} 
\end{eqnarray}
and
  \begin{eqnarray}
\frac{\partial \Omega}{\partial \bar{\Phi}} =\bigg[\frac{-a(T)\Phi}{2}-\frac{6b(T)
(\Phi-2{\bar{\Phi}}^2+{\Phi}^2\bar{\Phi})
}{\mathrm{1-6\bar{\Phi}\Phi+4(\bar{\Phi}^3+\Phi^3)-3(\bar{\Phi}\Phi)^2}}\bigg]T^4
-\sum_{i=u,d,s}\frac{2k_BTN_C}{(2\pi)^3}
\nonumber\\
\int_0^\infty d^3k\ \bigg[\frac{e^{-2(E_i^*(k)-{\nu_i}^{*})/k_BT}}{1+e^{-3(E_i^*(k)-{\nu_i}^{*})/k_BT}+3\Phi e^{-(E_i^*(k)-{\nu_i}^{*})/k_BT}
+3\bar{\Phi}e^{-2(E_i^*(k)-{\nu_i}^{*})/k_BT}}
\nonumber\\
+\frac{e^{-(E_i^*(k)+{\nu_i}^{*})/k_BT}}{1+e^{-3(E_i^*(k)+{\nu_i}^{*})/k_BT}
+3\bar{\Phi} e^{-(E_i^*(k)+{\nu_i}^{*})/k_BT}+3\Phi e^{-2(E_i^*(k)+{\nu_i}^{*})/k_BT}}\bigg]=0. 
\label{Polyakov conjugate} 
\end{eqnarray}

 In above, $m_\pi$ and $m_K$ denote the masses of $\pi$ and $K$ meson. The free model parameters $k_i(i=0,...,4)$, $g_s$, $g_v$, $g_4$, $h_1$ and $h_2$ can be calculated using $\pi$-meson mass, $K$-meson mass, the vacuum masses of $\sigma$, $\zeta$ and $\chi$ mesons and the average masses of $\eta$ and $\eta^{'}$  mesons \cite{wang2003}. Individual parameters used in this model  are listed in Table \ref{tab:2}.

\begin{table}[b]
\centering
\begin{tabular}{|c|c|c|c|c|c|c|c|c|c|}
\hline
$k_0$           & $k_1$          & $k_2$          & $k_3$         & $k_4$         & $g_s$         & $\rm{g_v}$          & $\rm{g_4}$           & $h_1$          & $h_2$                            \\ \hline
4.94                 & 2.12                & -10.16              & -5.38              & -0.06              & 4.76               & 10.92               & 37.5                 & -2.20               & 3.24                                  \\ \hline
$\sigma_0$ (MeV) & $\zeta_0$(MeV)  & $\chi_0$(MeV)   & $m_\pi$(MeV)  & $f_\pi$(MeV)  & $m_K$(MeV)    & $f_K$(MeV)     & $m_\omega$(MeV) & $m_\phi$(MeV)  & $m_\rho$( MeV)                   \\ \hline
-93                  & -96.87              & 254.6               & 139                & 93                 & 496                & 115                 & 783                  & 1020                & 783                                   \\ \hline
$g_{\sigma}^u$  & $g_{\sigma}^d$ & $g_{\sigma}^s$ & $g_{\zeta}^u$ & $g_{\zeta}^d$ & $g_{\zeta}^s$ & $g_{\delta}^u$ & $g_{\delta}^d$  & $g_{\delta}^s$ & $\rho_0$(fm$^{-3}$) \\ \hline
3.36                 & 3.36                & 0                   & 0                  & 0                  & 4.76               & 3.36                & -3.36                & 0                   & 0.15                                  \\ \hline
$g^u_{\omega}$  & $g^d_{\omega}$ & $g^s_{\omega}$ & $g^u_{\phi}$  & $g^d_{\phi}$  & $g^s_{\phi}$  & $g^u_{\rho}$   & $g^d_{\rho}$    & $g^s_{\rho}$   & $d$                              \\ \hline
3.86                 & 3.86                & 0                   & 0                  & 0                  & 5.46               & 3.86                & -3.86                & 0                   & 0.18                                  \\ \hline
\end{tabular}
\caption{The list of parameters used in the present work.}
\label{tab:2}
\end{table}

With the help of thermodynamical potential density, $\Omega$, one can calculate the pressure, $p$, free energy density, $F$, entropy density, $S$ and the energy density, $\epsilon$ using relations
\begin{equation}
p=-\Omega,
\label{p1}
\end{equation}
\begin{equation}
F=\Omega+\sum_{i=u, d, s} {\nu_i}^{*} {\rho_i},
\label{freeenergy}
\end{equation}
\begin{equation}
S=-\frac{\partial \Omega}{\partial T},
\end{equation}
and
\begin{equation}
\epsilon=\Omega+\sum_{i=u, d, s} {\nu_i}^{*} \rho_i+TS,
\label{energy1}
\end{equation}
respectively.

Strange quark matter produced in heavy ion collisions is metastable. The number of constituent quarks can be generally found unequal in HICs and therefore, isospin asymmetry can be incorporated through definition 
\begin{equation}
\eta=\frac{(\rho_d-\rho_u)}{(\rho_d+\rho_u)/3}.
\end{equation}
For asymmetric quark matter, the total baryon density can be expressed in terms of number density of quarks as $\rho_B=\frac{1}{3} (\rho_u+\rho_d+\rho_s)$, and the baryon, isospin and strangeness chemical potential are defined through
 $\mu_B=\frac{3}{2}(\mu_u+\mu_d)$, $\mu_I=\frac{1}{2}(\mu_u-\mu_d)$ and $\mu_S=\frac{1}{2}(\mu_u+\mu_d-2\mu_s)$, respectively. To characterize the flavor composition, we also introduce the strangeness fraction parameter, $f_{s}$=$\rho_{s}/\rho_B$.

\section{Numerical Results and Discussions}
\label{sec:4}
Now we will present the results on various thermodynamical properties of strange quark matter using Polyakov extended Chiral $\text{SU(3)}$ quark mean-field model. 
In the PCQMF model, the effect of temperature comes into picture through the  scalar densities, ${{\rho_{i}}^s}$ and vector densities, $\rho_i$ of constituent quarks which in turn depend upon scalar, vector and Polyakov loop fields.  As said before, the scalar fields  ($\sigma$, $\zeta$, $\delta$, $\chi$), the vector fields ($\omega$, $\rho$, $\phi$), and the Polyakov loop field, ($\Phi$, $\bar{\Phi}$) are calculated by solving the coupled system of non-linear equations (from eqs. (\ref{sigma1}) to (\ref{Polyakov conjugate})). In \cref{subsec_4a}, 
we will discuss the in-medium behavior of scalar, vector and Polyakov loop fields, which will be used as input to understand different thermodynamic properties of strange quark matter which are presented in \cref{subsec_4b}.  
\subsection{In-medium Scalar, Vector and Polyakov Loop Fields} 
\label{subsec_4a}
\begin{figure}
\includegraphics[width=19cm,height=21cm]{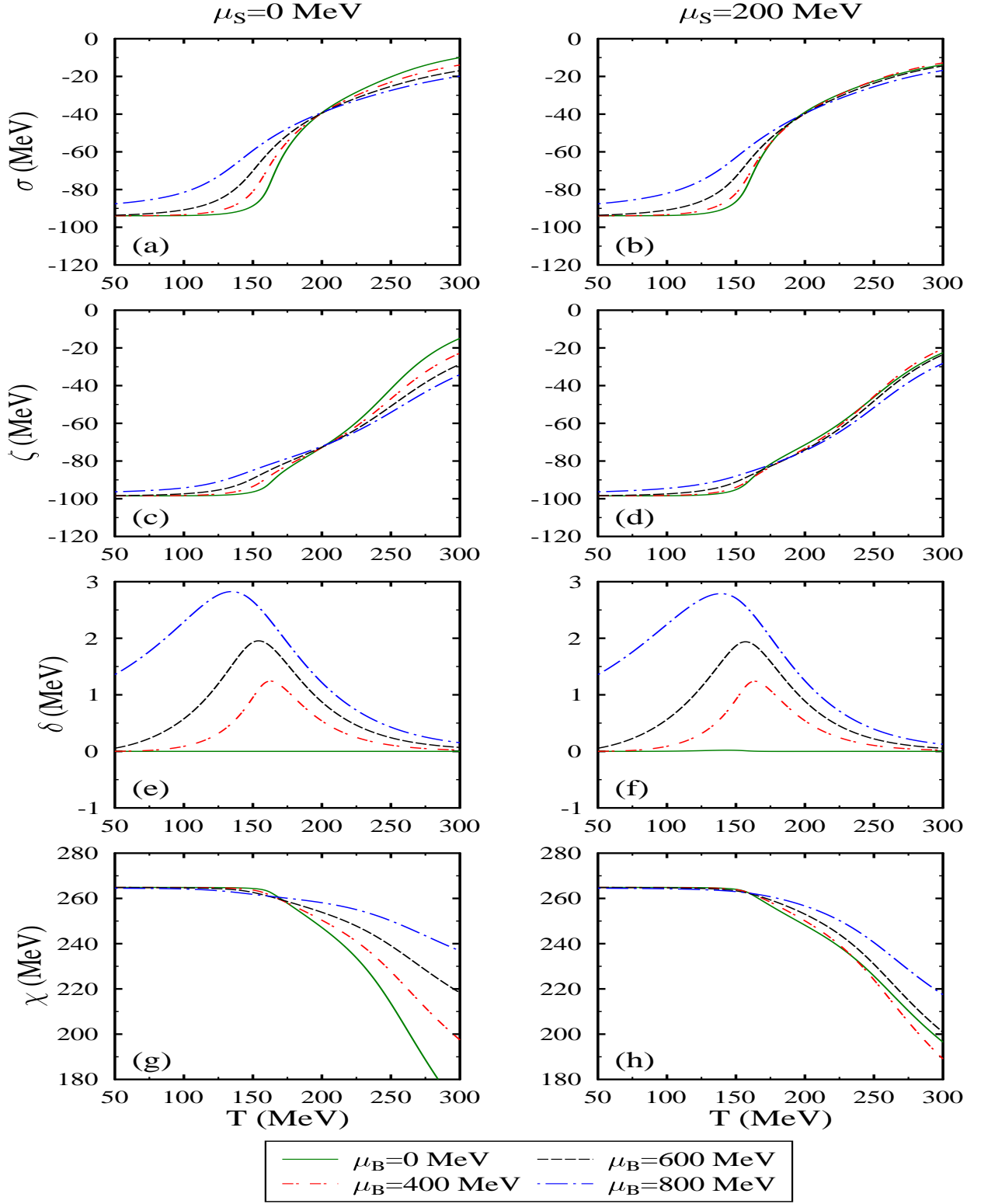}
\caption{(Color online) The scalar fields $\sigma$, $\zeta$, $\delta$ and $\chi$ plotted as a function of temperature $T$, for baryon chemical potential, $\mu_B=0$, 400, 600 and 800 MeV, strangeness chemical potential, $\mu_S=0$, 200 MeV and isospin chemical potential, $\mu_I=80$ MeV.}
\label{sfields}
\end{figure}

In \cref{sfields} we have shown the variation of $\sigma$, $\zeta$, $\delta$ and $\chi$ fields as a function of temperature, $T$, for strangeness chemical potential, $\mu_S = 0$ and 200 MeV and baryon chemical potential, $\mu_B = 0,$ 400, 600 and 800 MeV. The value of  isospin chemical potential is kept fixed at $\mu_I=80$ MeV.
As can be seen from \cref{sfields} (a) and (b), at $\mu_B = 0$, the magnitude of $\sigma$ and $\zeta$ remain constant upto a certain temperature and then start decreasing with further increase in temperature.
The temperature at which the magnitude of scalar fields starts decreasing sharply is named as pseudo-critical temperature $T_p$. Here, one can see that the value of $T_p$ decreases with an increase in baryon chemical potential $\mu_B$.
The large decrease in the magnitude of scalar fields at high value of temperature and/or baryon chemical potential may be a signature of restoration of chiral symmetry.
 With the increase in temperature, the fermi distribution function given in \cref{distribution,distribution1} decreases, which decreases the scalar density written in eq. (\ref{rhos0}). The smooth decrease in fields describes a slow transition (crossover) \cite{Schaefer2007}.
 For a given temperature, the magnitude of $\sigma$ and $\zeta$ fields decrease with an increase in $\mu_B$.
At $\mu_B$ = 800 MeV, the  magnitude of $\sigma$ and $\zeta$ drop significantly even at small temperature.
This favours the conclusion that the transition to a phase of quark matter can be achieved at the high baryonic density and moderate temperature such as in future CBM experiments of the FAIR project.
 
For a given temperature, at $\mu_B = 0$, the increase in the strangeness chemical potential $\mu_S$ causes a decrease in the magnitude of scalar fields $\sigma$ and $\zeta$. 
For example, at $T$ = 200 MeV and  $\mu_S=0$ (200) MeV, the magnitude of $\sigma$ and $\zeta$ fields are observed to be -39.23 (-38.98) and -72.79 (-71.41) MeV, respectively.  
This indicates that the increase in strangeness chemical potential leads to decrease in transition temperature (where the transition to the quark phase will take place).
On the other hand, for finite $\mu_B$, the magnitude of scalar fields increases with increase in $\mu_S$. For example, at $\mu_B = 800$ MeV and  $\mu_S = 0$ (200) MeV, the magnitude of $\sigma$ and $\zeta$  fields are calculated as -39.20 (-39.58) and -72.20 (-74.56) MeV.   
 At higher temperature, the effect of $\mu_B$ on magnitude of $\sigma$ and $\zeta$ fields is negligible for finite strangeness chemical potential.   
 
The scalar isovector field $\delta$ contributes to isospin asymmetry of the medium and is plotted in \cref{sfields}(e) and (f). It is observed that the magnitude of $\delta$ is zero for different values of temperature at $\mu_B=0$. This is because the $\delta$ field is calculated by the difference of $u$ and $d$ quarks scalar density and both of these are equal at zero baryon chemical potential.
 For a given temperature, the increase in the value of $\mu_B$ causes an increase in the value of $\delta$ field.
 This increase in the $\delta$ field as a function of $\mu_B$ is more in the low temperature regime.
 For finite $\mu_B$, the magnitude of $\delta$ increases with temperature up to a certain value and then starts decreasing with further increment in temperature.

  The trace anomaly property of QCD, which connect the trace of energy-momentum tensor to the expectation value of scalar gluon condensates is simulated in the chiral quark mean field through the scalar dilaton field $\chi$ which is plotted in \cref{sfields}(g) and (h). For temperature below $T_p$, the dilaton field remains almost constant even on varying $\mu_B$, which means a minute change is observed in the low temperature region. However, above $T_p$, the magnitude of $\chi$ decreases with an increase of $T$. For a given temperature (above $T_P$), the increase in baryon chemical potential is found to cause an increase in the dilaton field. When we consider the finite value of $\mu_S$, we  observe less change in the value of $\chi$ as compared to $\mu_S = 0$ at high temperature.  At temperature $T=200$ MeV and $\mu_S=0$, the value of $\chi$ field, at $\mu_B=0$, 400 and 800 MeV are observed to be 247.13, 250.37 and 258.10 MeV, respectively, whereas at $\mu_S=200$ MeV, above values changes to 248.15, 250.09 and 256.65 MeV.

\begin{figure}[ht]
\includegraphics[width=18cm,height=16cm]{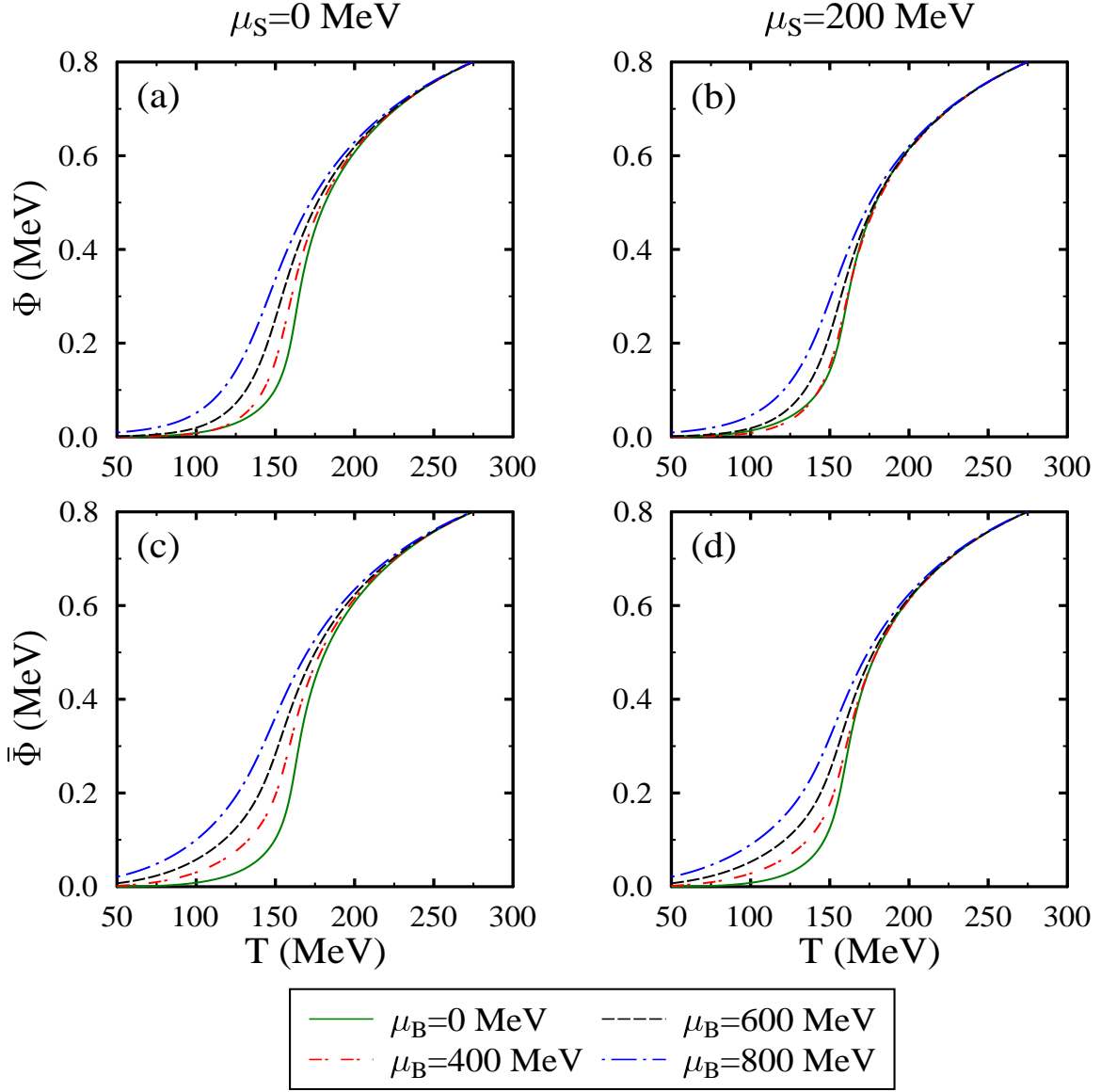}
\caption{(Color online) The Polyakov loop fields $\Phi$, and  $\bar{\Phi}$ plotted as a function of temperature $T$, for different values of strangeness chemical potential  $\mu_S$ with baryon chemical potential, $\mu_B=0$, 400, 600 and 800 MeV at isospin chemical potential $\mu_I=80$ MeV.}
 \label{pfields}
\end{figure}

  \begin{figure}[ht]
\includegraphics[width=18cm,height=16cm]{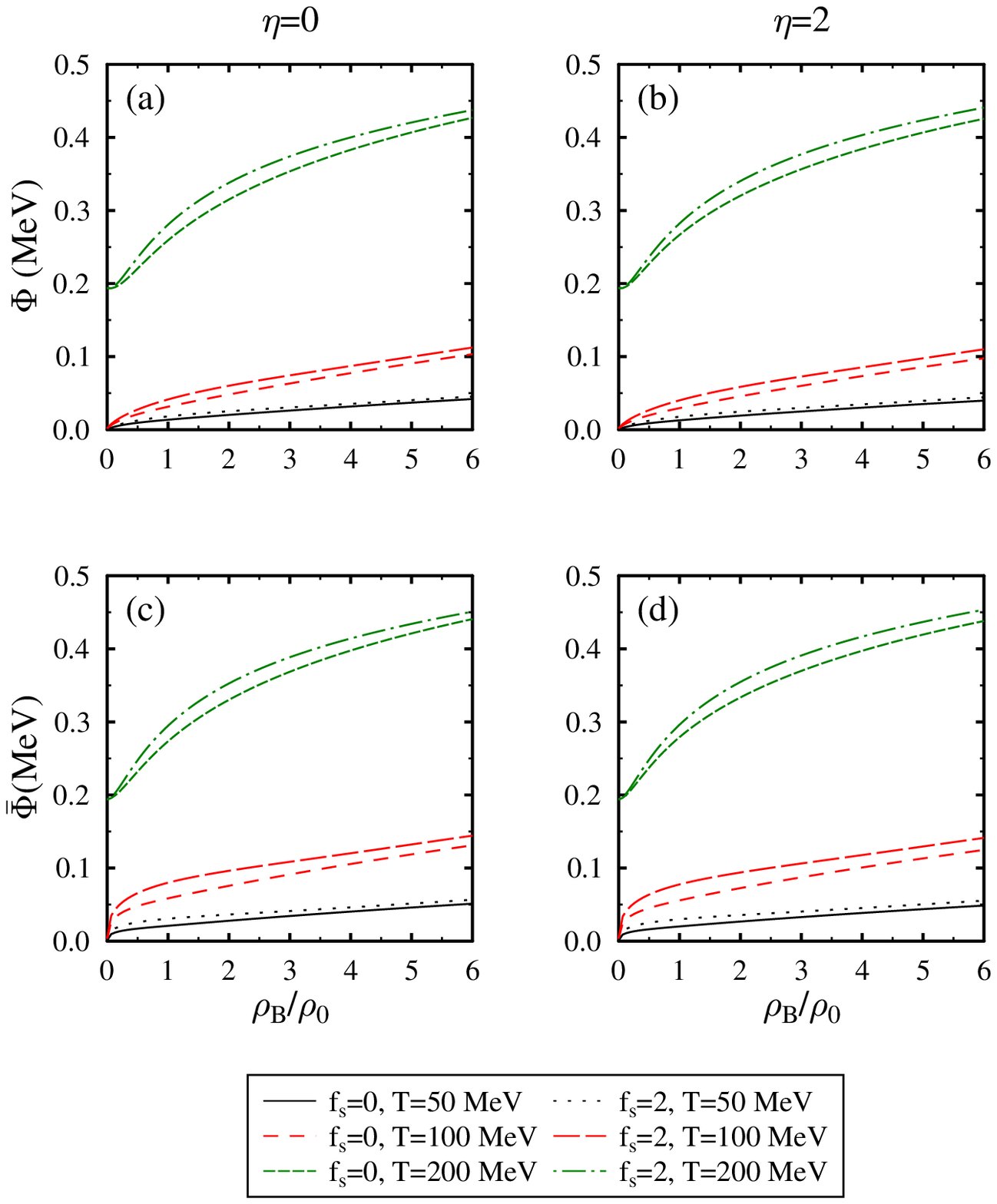}
\caption{(Color online) The Polyakov loop fields, $\Phi$ and  $\bar{\Phi}$ plotted as a function of baryon density, $\rho_B/\rho_0$ (in unit of nuclear saturation density), for different values of strangeness fraction and isospin asymmetry at $T$= 50, 100 and 200 MeV.}
 \label{Phi}
\end{figure}

\Cref{pfields} depicts the Polyakov loop fields $\Phi$ and $\bar{\Phi}$, known as a deconfinement order parameter, in the mean-field approximation.
 We found that at vanishing baryon chemical potential  ($\mu_B=0$ MeV) both $\Phi$ and $\bar{\Phi}$ are identical. The thermal effects of the confined state on the evolution of $\Phi$ looks to be very smooth. At low temperature, the slope of $\Phi$ and $\bar{\Phi}$ appears to depend on the temperature. The value of $\Phi$ and $\bar{\Phi}$ is approximately zero at lower temperature indicating that the system is in a confined state. With increase in temperature, $\Phi$ increases and at a higher value of temperature, the system converts from confined to the deconfined state. The increase of the baryon chemical potential, further causes an increase in the values of Polyakov loop fields. This may indicate a decrease in the deconfinement temperature.

 In order to understand more about the behavior of  $\Phi$ and  $\bar{\Phi}$, in \cref{Phi} we have plotted the variation of these fields as a function of baryon density, $\rho_B$ (in the unit of nuclear saturation density). The results are shown at zero and non-zero values of isospin asymmetry and strangeness fraction of the medium. 
The values of  $\Phi$ and  $\bar{\Phi}$ increases with increase in the density of the medium. Further, for given density, the increase in temperature also causes an increase in the values of these fields.  For a given density and temperature, as we move from $f_s$ = 0 to a finite value, further increment in the field is observed.
 
 \subsection{ Various Thermodynamical Quantities}
 \label{subsec_4b}
 
 \begin{figure}[ht]
\includegraphics[width=18cm,height=21cm]{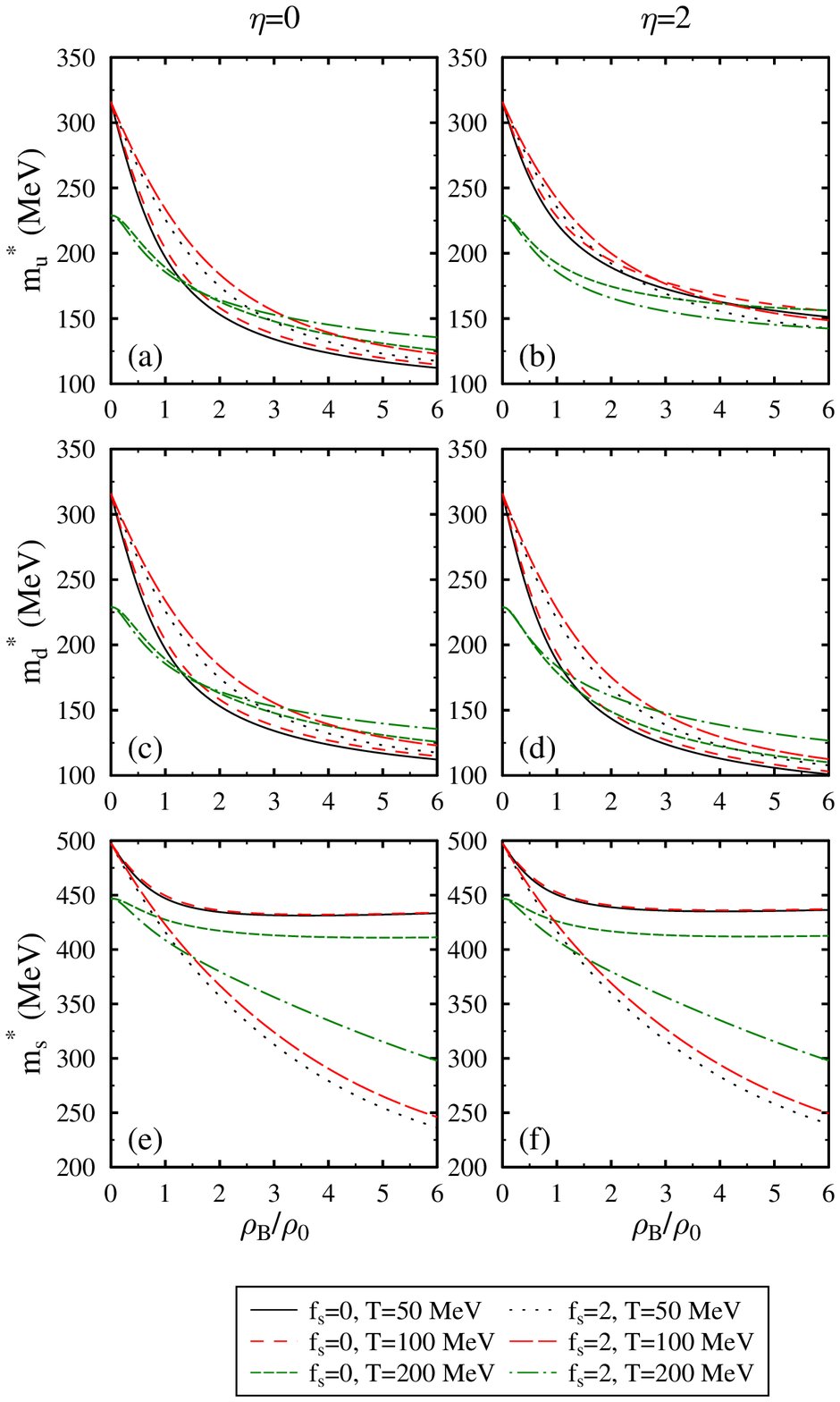}
\caption{(Color online) The effective constituent quark masses at $T=50$, 100 and 200 MeV, for different value of  $f_s$ and $\eta$, as a function of baryonic density, $\rho_B$ (in units of nuclear
saturation density $\rho_0$).}
 \label{mass}
\end{figure} 
 
   \begin{figure}[hb]
\includegraphics[width=18cm,height=10cm]{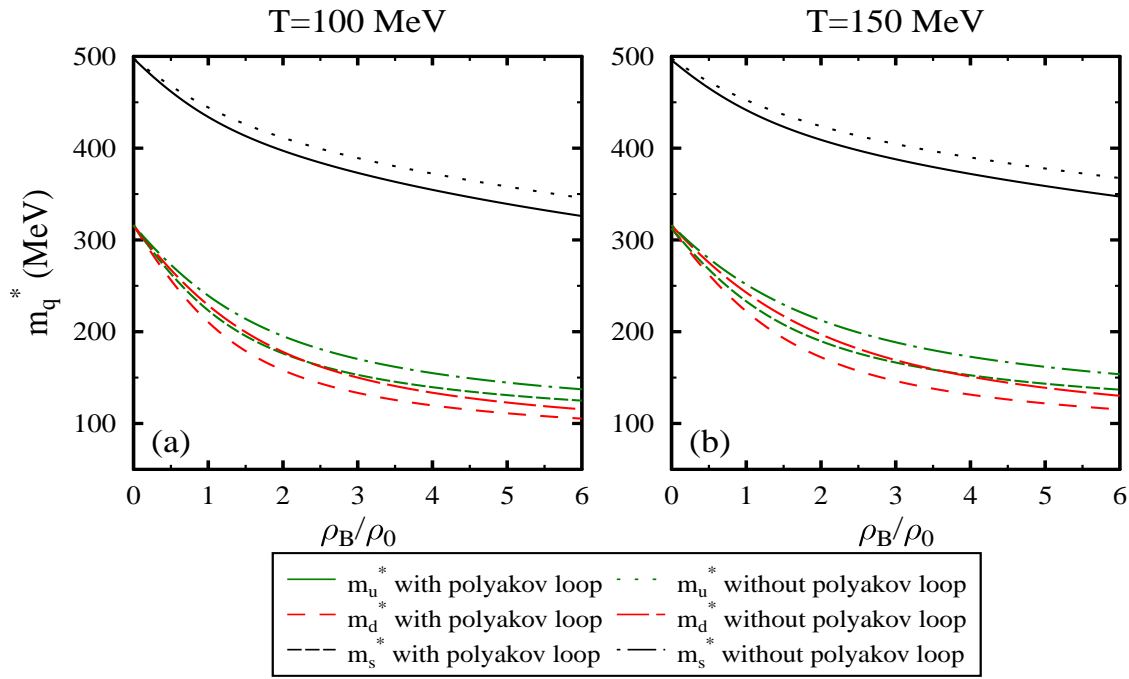}
\caption{(Color online) Comparison of effective masses of quarks at $T=100$ and 150 MeV for CQMF and PCQMF model as a function of $\rho_B/\rho_0$.}
 \label{mass_com}
\end{figure} 
  
 \begin{figure}[ht]
\includegraphics[width=18cm,height=11cm]{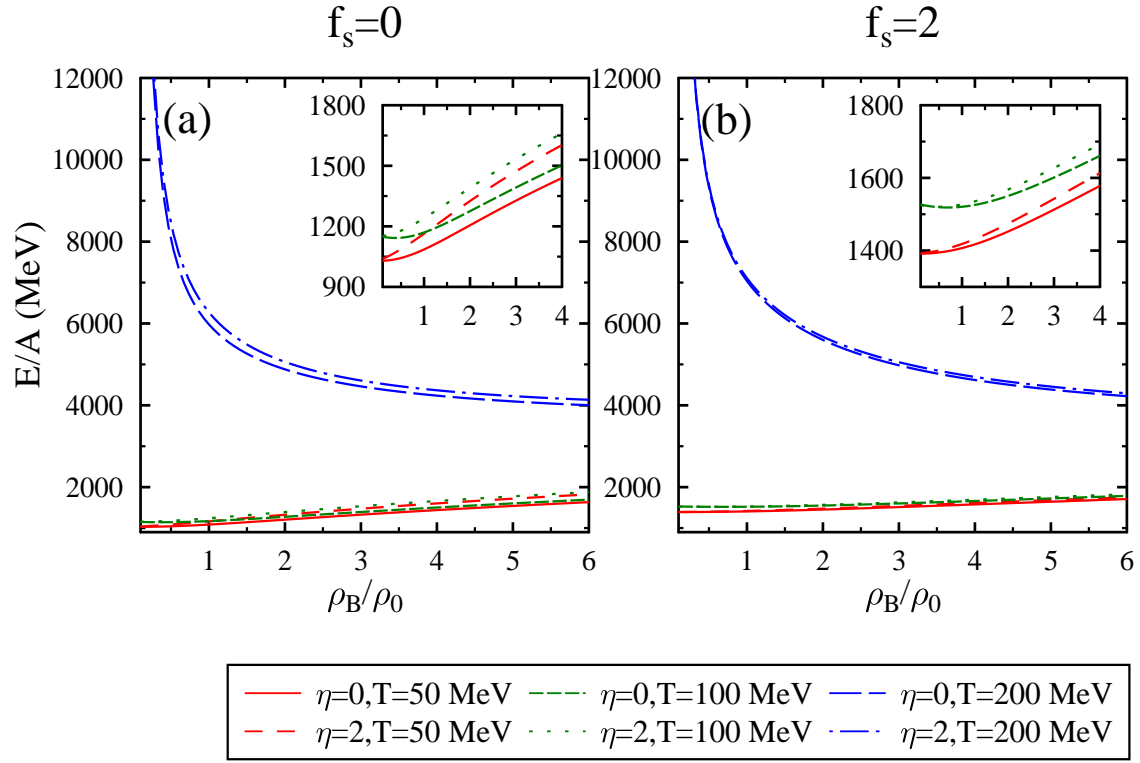}
\caption{(Color online) The energy per baryon of asymmetric strange quark matter at $T=50$, 100 and 200 MeV with baryonic density, $\rho_B$ (in units of nuclear saturation density $\rho_0$).}
 \label{epb}
\end{figure}

\begin{figure}[ht]
\includegraphics[width=18cm,height=11cm]{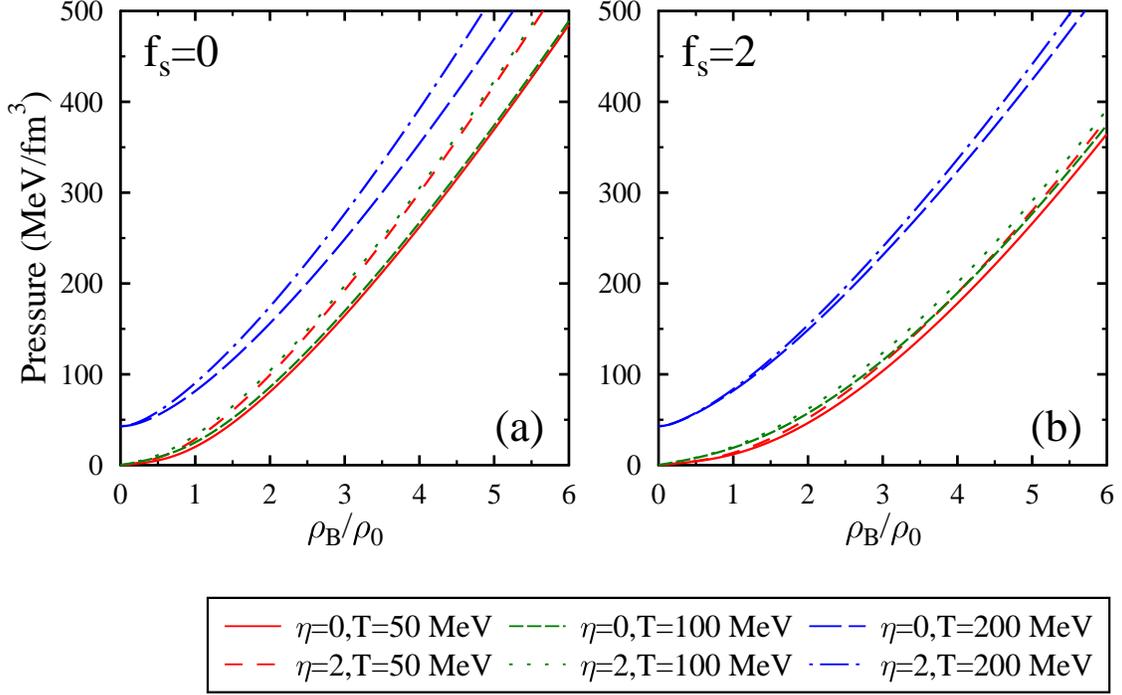}
\caption{(Color online) The behavior of pressure at $\eta=0$, 2 and $f_s=0$, 2 for different value of temperature as a function of baryonic density, $\rho_B$ (in units of nuclear saturation density $\rho_0$).}
 \label{pressure}
\end{figure}

\begin{figure}[ht]
\includegraphics[width=18cm,height=11cm]{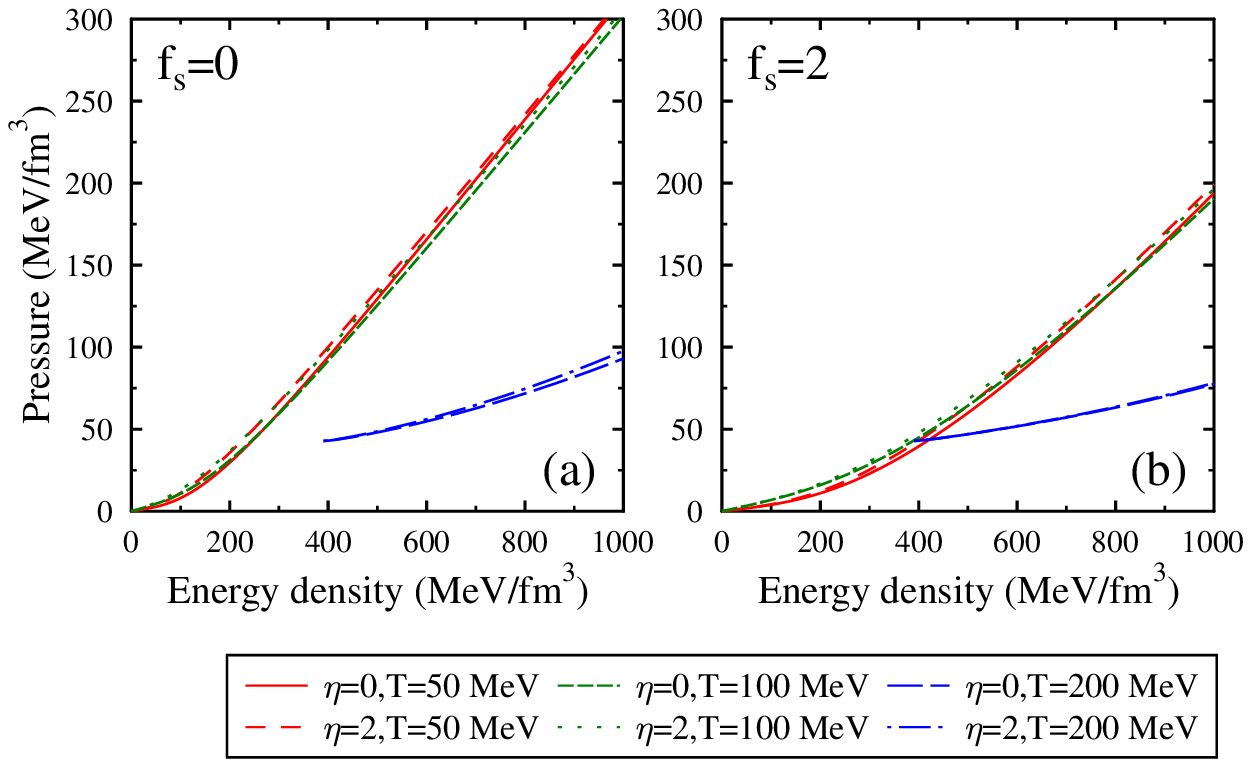}
\caption{(Color online) The EoS plotted as a function of baryonic density$\rho_B$ (in units of nuclear saturation density $\rho_0$), for different value of temperature, $T$, isospin asymmetry, $\eta$ and strangeness fraction, $f_s$.}
 \label{eos}
\end{figure}

The constituent quark masses are generated by the coupling of scalar fields  $\sigma$, $\zeta$ and $\delta$ with the quarks. In  \cref{mass}, we plot the quark masses, ${m_i}^{*}$ ($i = $ $u$, $d$ and $s$) as a function of total baryon density for temperatures $T$= 50, 100 and 200 MeV, for different value of ${f_s}$ and $\eta$.
For a given temperature, isospin asymmetry and strangeness fraction, the effective mass  ${m_i}^{*}$ of quarks are observed to decrease with an increase in the density of  medium. In a non-strange medium, the effective mass of $u$ and $d$ quarks
decreases more sharply as compared to the mass of strange $s$ quark. This result is due to zero value of coupling of $s$ quark with scalar $\sigma$ field ($g_\sigma^{s} = 0$). However, at finite strangeness, the effective mass of strange quarks shows
a rapid decrease with an increase in density. The reason is, at finite strangeness fraction,  the attractive interactions of strange quark with strange scalar field $\zeta$ dominate over the interaction of light $u$ and $d$ quarks with $\sigma$ field. 
  As can be seen from  \cref{mass}(a) and (c), in symmetric quark matter,  for a given density, at temperature $T$ = 50 and $100$ MeV 
  the value of $m_u^*$ increase  on moving from zero to finite value of $f_s$.  However, at very high temperature say $T = 200$ MeV, the trend becomes opposite at high baryonic density.  
  The isospin asymmetry of the medium causes the mass splitting between $u$ and $d$ quarks. As we change $\eta$ from zero to finite value, for fix value of other parameters, the in-medium mass of $u$ quark increases whereas the mass of $d$ quark decreases.   From  \cref{mass}(b), we observed that for finite $\eta$, at high baryonic density, $m_u^{*}$ becomes less in the strange medium as compared to the non-strange medium. This is opposite to $\eta$ = 0 situation as discussed above.

The strangeness and density dependence of the effective quark masses are evaluated in the $\text {SU(3)}$ NJL model \cite{Mishustin2001} and CQMF model at zero temperature \cite{wang2003} and observed that at large baryon density, the value of ${m_s}^{*}$ is even lower than the  ${m_u}^{*}$ (=${m_d}^{*}$).
   On comparing the result of PCQMF and CQMF at finite baryon density (in \cref{mass_com}), a larger value of quark mass is achieved in the presence of Polyakov loop potential at $T=100$ and 150 MeV. Furthermore, quark masses increases with an increase in the temperature at finite baryon density.

In \cref{epb}, we have depicted the energy per baryon $E/A$ as a function of total baryon density, $\rho_B$  within the PCQMF model for different values of strangeness fraction and isospin asymmetry, at $T$=50, 100, and 200 MeV. For temperatures $T = 50$ and $100$ MeV, the energy per baryon, $E/A$, increases uniformly with increasing $\rho_B$.
In Ref. \cite{wang2003},  energy per baryon was calculated for different values of vector coupling constant at zero temperature without Polyakov loop effect. In the present work, we have included\ the vector interaction at finite temperature and observe 
that $E/A$ increases with an increase in the temperature which is consistent with the calculations within equiparticle model \cite{Zhen2016}.
For fix value of density and temperature, an increase in the isospin asymmetry, $\eta$, or strangeness fraction $f_s$ of the medium also observed to cause an increase in the energy per baryon. 
For zero temperature, the minima of free energy coincide explicitly to the zero pressure \cite{peng2018}. Nonetheless, two points (minimum and zero pressure) of energy per baryon for SQM have not coincided at finite temperature (see  \cref{freeenergy,energy1}). 
With an increase in temperature, the position of minimum shifts from low  to high baryonic density.

\Cref{pressure} displays the pressure density of quarks with the total baryon density for SQM  at $T$ = 50, 100 and 200 MeV within the PCQMF model. We have observed that the pressure density increases gradually with the increase in $\rho_B$ for different values of temperature. For given density, isospin asymmetry and strangeness fraction, an increase of temperature cause an increase of pressure.
The isospin asymmetry of the medium also causes an increase in the value of pressure. The pressure value is more in case of non-strange medium when compared with the strange medium for both $\eta=0$ and 2.
 Considering the effect of strangeness fraction, the value of pressure decrease as a function of $f_s$ at a given temperature and density of the medium. Quark matter at finite temperature has also been studied in $\text SU(3)$ NJL model with different temperatures and fixed $f_s$ without considering vector coupling, which is close to our results \cite{Mishustin2000}. 
   
   In \cref{eos}, we have plotted the EoS, i.e., pressure density versus energy density at $T$ = 50, 100 and 200 MeV. The pressure density, $p$, is non-negative, smooth, and monotonically increasing with energy density. We have perceived that EoS gets stiffer when the temperature increases in non-strange as well as in strange medium, both in symmetric and asymmetric quark matter. 
   For a given temperature and density, an increase of asymmetry in the medium also cause an increase in the stiffness of the EoS. However, an increase in strangeness fraction, for fix value of other parameters, causes softness in the EoS.
   The EoS of quark matter has also been explained in Ref. \cite{peng2017} using the NJL model for various interactions at finite temperature. It was observed that the repulsive vector interaction $G_V$ and $g_V$ both contributed equally and stiffen the equation of state.

   \begin{figure}[ht]
\includegraphics[width=18cm,height=16cm]{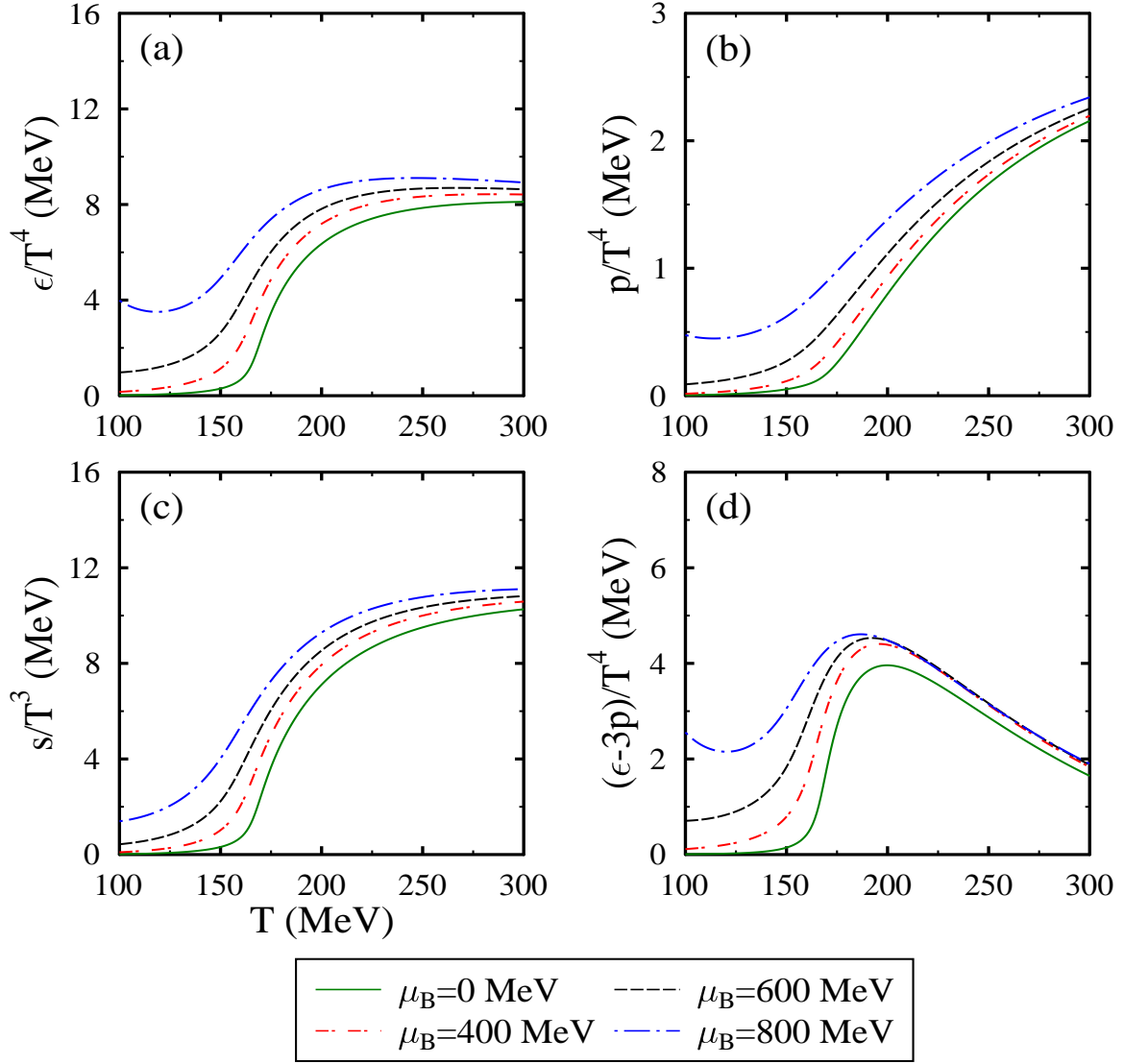}
\caption{(Color online) The energy densty, $\epsilon$, pressure density, $p$, entropy density, $s$ and trace anomaly, $(\epsilon-3p)/T^4$ as a function of temperature $T$, for two flavor quark matter ($u$ and $d$ quarks) with baryon chemical potential, $\mu_B=0$, 400, 600 and 800 MeV at isospin chemical potential $\mu_I=80$ MeV.}
 \label{energy2f}
\end{figure} 

  The energy density and pressure calculated within the PCQMF model can be utilized further to calculate the quantity
   $(\epsilon-3p)/T^4$ representing the trace anomaly property of QCD.
In \cref{energy2f}, we have displayed the temperature dependence of  $p/T^4$, $\epsilon/T^4$, $s/T^3$ and $(\epsilon-3p)/T^4$ with baryon chemical potential, $\mu_B=0$, 400, 600 and 800 MeV at isospin chemical potential $\mu_I=80$ MeV for two flavor quark matter ($u$ and $d$). The energy density, pressure and entropy  are continuous functions of temperature showing that the transition to the QGP phase is a crossover instead of a phase transition.
 If we increase $\mu_B$, the magnitude of all thermodynamical quantities increases smoothly for a constant temperature. Also, we have observed a comparable behavior in a $p/T^4$, $\epsilon/T^4$ and $s/T^3$ curve $i.e.$  a sharp increase near the transition temperature and then approach to corresponding ideal gas limit. 
 In QCD, asymptotic properties imply that the trace anomaly becomes dependent on the strength of strong coupling constant $((\epsilon-3p)/T^4 \propto T^4{\alpha_s}^2$) \cite{Tawfik2013}. The trace anomaly is vanishing or very small for the confined phase in scale-invariant theory. At $T$ $\ll$ $T_p$, the trace anomaly is sufficiently small and then increases with an increase in temperature. This is described by large interaction strength, which brings quarks and gluons close and then they bound in hadrons. For $T$ $\gg$ $T_p$, the interaction strength becomes weaker and weaker. This implies that the quarks and gluons become free, especially at very high $T$, where $(\epsilon-3p)/T^4\approx 0$. The smooth crossover could be identified by a peak of $(\epsilon-3p)/T^4$. 
 \begin{figure}[ht]
\includegraphics[width=18cm,height=20cm]{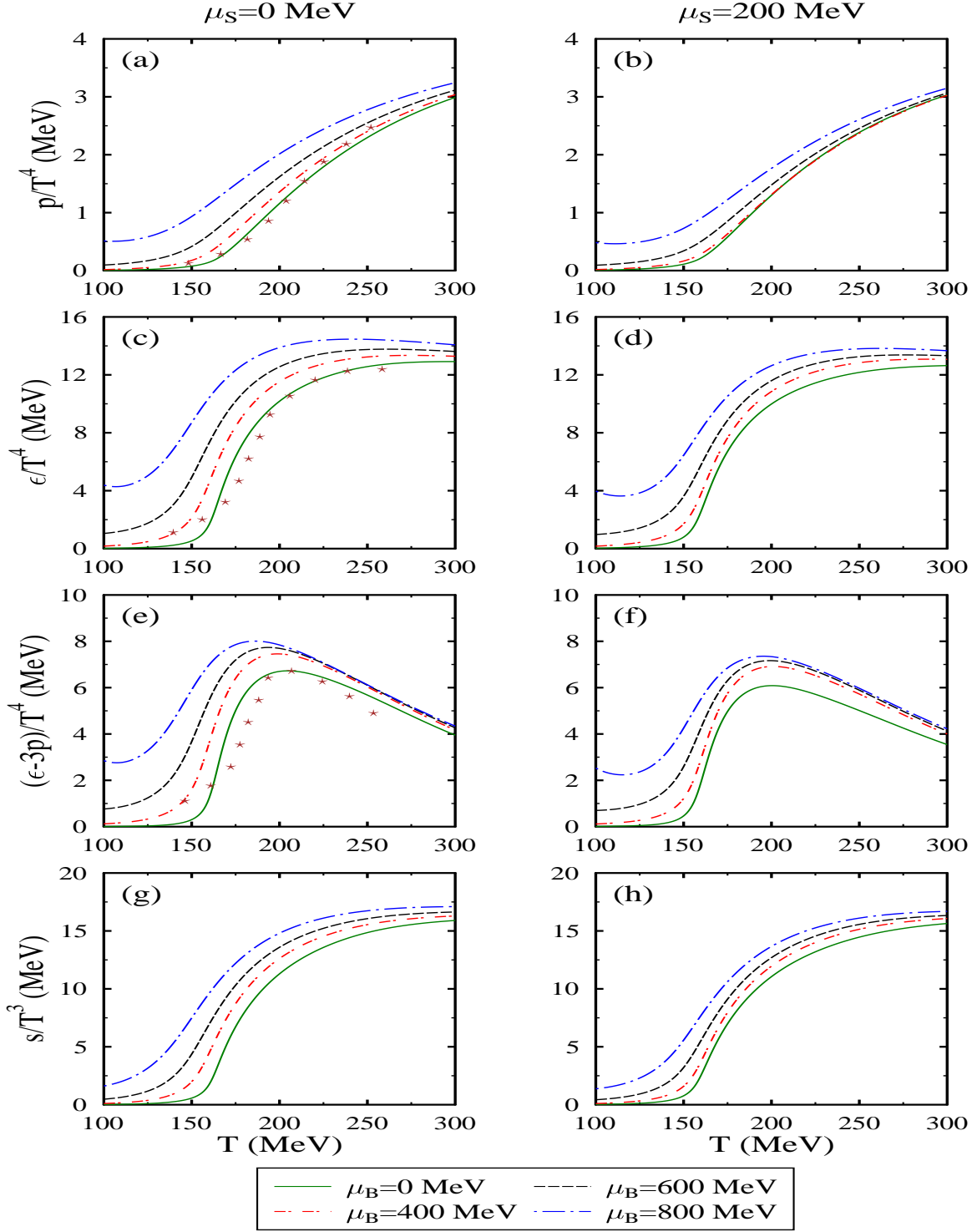}
\caption{(Color online) The energy densty, $\epsilon$, pressure density, $p$, entropy density, $s$ and trace anomaly, $(\epsilon-3p)/T^4$ as a function of temperature $T$, for different values of strangeness chemical potential  $\mu_S$ with baryon chemical potential, $\mu_B=0$, 400, 600 and 800 MeV at isospin chemical potential $\mu_I=80$ MeV, for three flavor quark matter ($u$, $d$ and $s$ quarks).}
 \label{energy3f}
\end{figure} 

 In \cref{energy3f}, we have plotted the pressure density, the energy density, entropy density and the trace anomaly as a function of temperature with different baryon chemical potential and at $\mu_I=80$ for three flavor quarks ($u$, $d$ and $s$ quark) and compared with lattice QCD results \cite{Cheng2010}. 
 The Stefan Boltzmann (SB) limit of QCD also changes with increase in the number of flavors \cite{Costa2010,Borsanyi2012}. These quantities have also been studied using PNJL model \cite{Costa2010,Ratti2006} and PQM model \cite{Megias2006,Megias2006a,Abhishek2018} for two and three flavors. From \cref{energy2f,energy3f}, we observed that the values of pressure density and other thermodynamical quantities in the medium increase as one move from two to three flavor matter. An abrupt change in these quantities leads to deconfinement behavior with the formation of new degrees of freedom. The value of deconfinement temperature is larger in case of two flavors.  
 In both hadronic and QGP regions, a good agreement with recent lattice QCD calculations \cite{Cheng2010} is obtained. Calculation of bulk thermodynamic quantities for non-vanishing chemical potential based on the LQCD approach gives significant information about the phase structure of QCD \cite{Cheng2010}. Pressure and energy density of \text{SU(3)} gauge theory are calculated by performing simulations of (2+1)-flavor QCD using the highly improved staggered quarks (HISQ/tree) action on $N_\tau=4$, 6, 8 and 12 lattices and spatial extent $N_\sigma=16$ and 32  \cite{Bazavov2012,Bazavov2019,Bazavov2017,Ding2019,Ding2019a,
Ding2015,Borsanyi2010,Borsanyi2012}. These results are then extrapolated to the continuum limits.

\section{Summary}
To summarize, we have extended the CQMF model to include Polyakov fields and studied the properties of non-strange and strange quark matter at finite temperature and density. Within the PCQMF model, the effect of temperature through Polyakov loop potential and the scalar and vector densities of quarks  result in the modification of scalar, vector and Polyakov fields.
 The scalar and vector fields are further used to calculate effective constituent quark masses. The value of the constituent masses of $u$ and $d$ quarks become larger with strangeness fraction, whereas effective mass of $s$ quark decrease with $f_s$. In order to explore the properties of SQM, we have studied the energy density, pressure density, and EoS for different values of strangeness and isospin asymmetry. It is found that the EoS gets stiffer with increase in isospin asymmetry. We can also observed that the pressure density increases monotonically with baryon density at finite temperature, and enhanced when isospin asymmetry increase, which shows that the isospin effect contributes more with increment of temperature. We have further analysed the temperature dependence on the various thermodynamic quantity including pressure, entropy, energy density, and trace anomaly. The thermodynamical quantities of the PCQMF model such as $p/T^4$, $\epsilon/T^4$ and $(\epsilon-3p)/T^4$ are compared with the recent lattice QCD simulations. 
In future, we will focus on the study of $\beta$ equilibrated quark matter and magnetic field effects.

 \section*{Acknowledgment}

Authors sincerely acknowledge the support towards this work from the Ministry of Science and Human Resources (MHRD), Government of India via Institute fellowship under the National Institute of Technology Jalandhar. Arvind Kumar sincerely acknowledges the DST-SERB, Government of India for funding of research project CRG/2019/000096.


\begin{thebibliography}{99}

\bibitem{Boyanovsky2006}
D. Boyanovsky $et$  $al$., Ann. Rev. Nucl. Part. Sci. {\bf 56}, 441 (2006).
\bibitem{Boeckel2011}
 T. Boeckel $et$  $al$., Prog. Part. Nucl. Phys. {\bf 66}, 266 (2011).
\bibitem{Marty2013}
R. Marty  and J. Aichelin, Phys. Rev. C  {\bf 87}, 034912 (2013).
\bibitem{Janka2012}
H. T. Janka, Ann. Rev. Nucl. Part. Sci. {\bf 62}, 407 (2012).
\bibitem{Weber2005}
 F. Weber, Prog. Part. Nucl. Phys. {\bf 54}, 193 (2005).
 
 \bibitem{alford2001}
M. Alford, Ann. Rev. Nucl. Part. Sci. {\bf 51}, 131 (2001).

\bibitem{Rischke2004}
D.H. Rischke, Prog. Part. Nucl. Phys. {\bf 52},  197 (2004).
 
\bibitem{Fodor2004}
Z. Fodor and S.D. Katz, J. High Energy Phys. {\bf 04}, 050 (2004). 
 
 \bibitem{Iwasaki2004}
 Y. Iwasaki $et$  $al$., Phys. Rev. D {\bf 69}, 014507 (2004).
 
 \bibitem{McLerran2007}
 L. McLerran and R. D. Pisarski, Nucl. Phys. A {\bf 796}, 83 (2007).
 
 \bibitem{McLerran2009}
 L. McLerran, Nucl. Phys. A {\bf 830}, 709 (2009). 
 

 \bibitem{Fukushima2004}
 K. Fukushima, Phys. Lett. B {\bf 591}, 277 (2004).
 
\bibitem{Ratti2006}
 C. Ratti, M. A. Thaler and W. Weise, Phys. Rev. D {\bf 73}, 014019 (2006).
 
 \bibitem{Pisarski2000}
 R. D. Pisarski, Phys. Rev. D {\bf 62}, 111501 (2000).
 
 \bibitem{Fukushima2008}
 K. Fukushima, Phys. Rev. D {\bf 77}, 114028 (2008).

\bibitem{menezes2006}
D.P. Menezes $et$  $al$.,  J. Phys. G {\bf 32}, 1081 (2006). 

\bibitem{Kalam2013}
M. Kalam $et$  $al$.,  Int. J. Theor. Phys. {\bf 52}, 3319 (2013).

\bibitem{Lastowiecki2015}
R. Lastowiecki $et$  $al$., Phys. Part. Nuclei {\bf 46}, 843 (2015).

\bibitem{Annala2019}
E. Annala $et$  $al$., arXiv:1903.09121v1, astro-ph.HE (2019).

\bibitem{Bodmer1971}
 A. Bodmer, Phys. Rev. D {\bf 4}, 1601 (1971).
\bibitem{Farhi1984}
 E. Farhi and R. Jaffe, Phys. Rev. D {\bf 30}, 2379 (1984).
\bibitem{Witten1984}
 E. Witten, Phys. Rev. D {\bf 30}, 272 (1984).
 
\bibitem{Thirukkanesh2017}
S. Thirukkanesh and F.C. Ragel, Chin.Phys. C {\bf 41}, 015102 (2017). 

\bibitem{Shahzad2019}
M. R. Shahzad and G. Abbas, Int. J. Geom. Methods Mod. Phys. {\bf 16}, 1950132 (2019). 



\bibitem{ramona2007}
Ramona Vogt, \textquotedblleft Ultra-relativistic Heavy-Ion Collisions\textquotedblright ,   Elsevier (2007).
\bibitem{helios1995}
N. Masera, Nucl. Phys. A {\bf 590}, 93 (1995).
\bibitem{agaki1995}
G. Agakichiev $et$  $al$.,  Phys. Rev. Lett. {\bf 75}, 1272 (1995).
\bibitem{porter1997}
R. J. Porter $et$  $al$., Phys. Rev. Lett., {\bf 79}, 1229 (1997).
\bibitem{wilson1998}
W. K. Wilson $et$  $al$., Phys. Rev. C {\bf 57}, 1865 (1998).
\bibitem{sashi2011} 
Mahajan Gulshan, Shashi K. Dhiman, Phys. Rev. C {\bf 84}, 045804 (2011).
\bibitem{Hbook1994}
 C. Y. Wong, \textquotedblleft Introduction to High Energy Heavy Ion Collisions\textquotedblright, World Scientific Publishing Co. (1994).
\bibitem{sissa2009}
A. Sissakian  $et$  $al$., J.Phys. G {\bf 36}, 064069 (2009).
\bibitem{keke2012}
V. Kekelidze  $et$  $al$., Phys. Atom. Nucl. {\bf 75}, 542 (2012).

\bibitem{Roberts2000}
 C. D. Roberts and S. M. Schmidt,  Prog. Part. Nucl. Phys. {\bf 45}, S1 (2000).
 
 \bibitem{Alkofer2001}
 R. Alkofer and L. von Smekal,  Phys. Rep. {\bf 353}, 281 (2001).
 
  \bibitem{Maris2003}
 P. Maris and C. D. Roberts,  Int. J. Mod. Phys. E {\bf 12}, 297 (2003).
 
 
  \bibitem{Xu2015}
 S. S. Xu $et$  $al$., Phys. Rev. D {\bf 91}, 056003 (2015).
\bibitem{Fowler1981}
G.N. Fowler  $et$  $al$., Z. Phys. C {\bf 9}, 271 (1981).
\bibitem{Chakrabarty1989}
S. Chakrabarty $et$  $al$., Phys. Lett. B {\bf  229}, 112 (1989).
\bibitem{Chakrabarty1991}
S. Chakrabarty $et$  $al$., Phys. Rev. D {\bf  43}, 627 (1991).
\bibitem{Chakrabarty1993}
S. Chakrabarty $et$  $al$., Phys. Lett. D {\bf  48}, 1409 (1993).
\bibitem{Benvenuto1995}
O. G. Benvenuto and G. Lugones, Phys. Lett. D {\bf  51}, 1989 (1995).
\bibitem{Tsushima1998}
 K. Tsushima $et$  $al$., Nucl. Phys. A {\bf 630}, 691 (1998).
\bibitem{Schaefer2007}
 B. J. Schaefer $et$  $al$., Phys. Rev. D {\bf 76}, 074023 (2007).
 
\bibitem{Stiele2014}
 R. Stiele $et$  $al$., Phys. Lett. D {\bf 729}, 72 (2014).
 
\bibitem{peng2017}
Peng-Cheng Chu $et$  $al$., Eur. Phys. J. C {\bf 77}, 512 (2017).



\bibitem{Costa2010}
P. Costa $et$  $al$.,  Symmetry {\bf 2}, 1338 (2010).

\bibitem{Sakai1991}
Y. Sakai $et$  $al$., Phys. Rev. D {\bf  79}, 096001 (1991).

\bibitem{Sasaki1991}
T. Sasaki $et$  $al$., Phys. Rev. D {\bf  82}, 116004 (1991).

\bibitem{Sakai2010}
Y. Sakai $et$  $al$., Phys. Rev. D {\bf  82}, 076003 (2010).

\bibitem{Sasaki2012}
T. Sasaki $et$  $al$., Phys. Rev. D {\bf  85}, 056009 (2012).

\bibitem{Restrepo2015}
T. E. Restrepo $et$  $al$., Phys. Rev. D {\bf  91}, 065017 (2015).

\bibitem{Gatto2011}
 R. Gatto and M. Ruggieri, Phys. Rev. D {\bf  83},   034016 (2011).

\bibitem{Peng2000}
G.X. Peng $et$  $al$., Phys. Rev. C {\bf 62}, 025801 (2000).

\bibitem{Peng2008}
G.X. Peng $et$  $al$., Phys. Rev. C {\bf 77}, 065807 (2008).

\bibitem{Rajagopal2011}
K. Rajgopal, F. Wilczek Phys. Rev. Lett. {\bf 86}, 3492 (2011).

\bibitem{wang2001}
 P. Wang $et$  $al$., Commun. Theor. Phys. {\bf 36}, 71 (2001).

\bibitem{wang2001a}
P. Wang $et$  $al$., Nucl. Phys. A {\bf 688}, 791 (2001).

\bibitem{Chin1979}
S. Chin and A. Kerman, Phys. Rev. Lett. {\bf 43}, 1292 (1979).
\bibitem{Berger1987}
M. S. Berger and R. L. Jaffe, Phys. Rev. C {\bf 35}, 213 (1987).
\bibitem{Xu2008}
R. Xu, Mod. Phys. Lett. A {\bf 23}, 1629 (2008).



\bibitem{Zhang2001} 
Y. Zhang $et$ $al$., Europhys. Lett. {\bf 56}, 361 (2001). 
 
 \bibitem{Zhang2002} 
 Y. Zhang and R.K. Su, Phys. Rev. C {\bf 65}, 035202 (2002).

\bibitem{Qian2005} 
 W. L. Qian and R.K. Su, Int. J. Mod. Phys. A {\bf 20}, 1931 (2005).
 
 \bibitem{Bentz2001}
 W. Bentz and A. W. Thomas, Nucl. Phys. A {\bf 696} (2001).                            

\bibitem{wang2007}
P. Wang $et$ $al$., Phys. Rev. C {\bf 75}, 045202 (2007).

\bibitem{Schwarz1999}
T.M. Schwarz, S.P. Klevansky, and G. Rapp, Phys. Rev. C {\bf 60}, 055205 (1999).

\bibitem{Buballa1999}
M. Buballa ands M. Oertel, Phys. Lett. B {\bf 457}, 261 (1999). 

\bibitem{Liu2019}
H. Liu, J. Xu and C. M. Ko, Phys. Lett. B {\bf 803}, 135343 (2020).
 
 

\bibitem{Harpreet2018}
Harpreet Singh $et$  $al$., Eur. Phys. J. A  {\bf 54}, 120 (2018).



\bibitem{wang2002}
P. Wang $et$  $al$., Nucl. Phys. A {\bf 705}, 455 (2002).

\bibitem{wang2004a}
P. Wang $et$  $al$., Nucl. Phys. A {\bf 744}, 273 (2004).

\bibitem{wang2003}
 P. Wang $et$  $al$., Phys. Rev. C {\bf 67}, 015210 (2003).



\bibitem{Harpreet2018a}
Harpreet Singh $et$  $al$., arXiv:1811.05125v1, hep-th (2018).


%
%

\bibitem{Herbst2014}
Tina Katharina Herbst et.al., Phys. Lett. B {\bf 731}, 248 (2014).

\bibitem{Drews2013}
Matthias Drews et.al., Phys. Rev. D {\bf 88}, 09601 (2013).

\bibitem{Papazoglou1999}
 P. Papazoglou et.al., Phys. Rev. C {\bf 59}, 411 (1999).

\bibitem{Mishra2004}
A. Mishra et.al., Phys. Rev. C {\bf 69}, 015202 (2004).

\bibitem{Mishra2004a}
A. Mishra et.al., Phys. Rev. C {\bf 69},  024903 (2004).

\bibitem{Weinberg1968}
Steven Weinberg, Phys. Rev. {\bf 166}, 1568 (1968).
%
\bibitem{Coleman1969}
S. Coleman et.al., Phys. Rev. {\bf 177}, 2239 (1969).

\bibitem{Bardeen1969}
W. A. Bardeen and B. W. Lee, Phys. Rev. {\bf 177}, 2389 (1969).

\bibitem{shao2016}
G. Y. Shao$et$  $al$., Phys. Rev. D {\bf 94}, 014008 (2016).

\bibitem{Polyakov1978}
A. M. Polyakov, Phys. Lett. B {\bf 72}, 477 (1978).


\bibitem{Roessner2007}
Simon Roessner $et$  $al$., Phys. Rev. D {\bf 75}, 034007 (2007).

\bibitem{Fukugita1990}
M. Fukugita, M.  Okawa, and A. Ukava, Nucl. Phys. B {\bf 337}, 181 (1990).


\bibitem{Mishustin2001}
I. N. Mishustin et.al., Phys. At. Nucl. {\bf 64}, 802 (2001).



%



\bibitem{Zhen2016}
Zhen-Yan Lu $et$  $al$., Nucl. Sci. Tech. {\bf 27}, 148 (2016).

\bibitem{peng2018}
Peng-Cheng Chu $et$  $al$., Phys. Lett. C {\bf 778}, 447 (2018).

\bibitem{Mishustin2000}
I. N. Mishustin et.al., Phys. Rev. C {\bf 62}, 034901 (2000).

\bibitem{Tawfik2013}
A. Tawfik, Phys. Rev. C {\bf 88}, 035203 (2013).

\bibitem{Cheng2010}
M. Cheng $et$  $al$., Phys. Rev. D {\bf 81}, 054504 (2010).
%
%
%
%
%
%
%
%
%

%
\bibitem{Megias2006}
E. Megias $et$ $al$., Phys. Rev. D {\bf 74},  065005 (2006). 

\bibitem{Megias2006a}
E. Megias $et$ $al$., Phys. Rev. D {\bf 74},  114014 (2006). 

\bibitem{Abhishek2018}
A, Abhishek, H. Mishra and S. Ghosh, Phys. Rev. D {\bf 97}, 014005 (2018).

\bibitem{Bazavov2012}
A. Bazavov $et$ $al$., Phys. Rev. D {\bf 85},  054503 (2012). 

\bibitem{Bazavov2019}
A. Bazavov $et$ $al$., Phys. Rev. D {\bf 795},  15 (2019). 

\bibitem{Bazavov2017}
A. Bazavov $et$ $al$., Phys. Rev. D {\bf 95},  054504 (2017). 

\bibitem{Ding2019} 
H. T. Ding $et$ $al$., Phys. Rev. Lett. {\bf 123}, 062002 (2019).

\bibitem{Ding2019a} 
H. T. Ding $et$ $al$., Nucl. Phys. A {\bf 982}, 211 (2019).

\bibitem{Ding2015} 
H. T. Ding, F. Karsch and S. Mukherjee, Int. J. Mod. Phys. {\bf E24}, 1530007 (2015).

\bibitem{Borsanyi2010}
Sz. Borsanyi $et$ $al$., J. High Energy Phys.  {\bf 09}, 073 (2010).

\bibitem{Borsanyi2012}
Sz. Borsanyi $et$ $al$., J. High Energy Phys.  {\bf 08}, 053 (2012).

\end{thebibliography}

\end{document}